\documentclass[12pt, draftclsnofoot, onecolumn]{IEEEtran}

\usepackage{multicol}
\usepackage{amsmath,amssymb,mathtools,amscd,latexsym,dsfont,amsthm,bbold}
\usepackage[final]{graphicx}
\usepackage[usenames, dvipsnames]{color}
\usepackage{psfrag}
\usepackage{suffix}
\usepackage{cite}
\usepackage{fancyhdr}
\usepackage{mathrsfs}
 \usepackage{enumitem}

\makeatletter
\long\def\@makecaption#1#2{\ifx\@captype\@IEEEtablestring%
\footnotesize\begin{center}{\normalfont\footnotesize #1}\\
{\normalfont\footnotesize\scshape #2}\end{center}%
\@IEEEtablecaptionsepspace
\else
\@IEEEfigurecaptionsepspace
\setbox\@tempboxa\hbox{\normalfont\footnotesize {#1.}~~ #2}%
\ifdim \wd\@tempboxa >\hsize%
\setbox\@tempboxa\hbox{\normalfont\footnotesize {#1.}~ }%
\parbox[t]{\hsize}{\normalfont\footnotesize \noindent\unhbox\@tempboxa#2}%
\else
\hbox to\hsize{\normalfont\footnotesize\hfil\box\@tempboxa\hfil}\fi\fi}
\makeatother

\DeclareMathOperator{\erfc}{erfc}

\newcommand{\tr}[1]{\mathrm{#1}}

\DeclarePairedDelimiterX\MeijerM[3]%
{\lparen}{\rparen}{#3\,\delimsize\vert\ \begin{matrix}#1 \\ #2\end{matrix}}  
\newcommand\MeijerG[8][]{%
  \tr{G}^{#2,#3}_{#4,#5}\MeijerM[#1]{#6}{#7}{#8}}

\WithSuffix\newcommand\MeijerG*[7]{%
  \tr{G}^{#1,#2}_{#3,#4}\MeijerM*{#5}{#6}{#7}}

  \DeclarePairedDelimiterX\MultiMeijerM[2]%
  {\lparen}{\rparen}{#1\delimsize\vert\ #2}  
 
  \WithSuffix\newcommand\MultiMeijerG*[4]{%
    \tr{G}^{#1}_{#2}\MultiMeijerM*{#3}{#4}}
\newcommand\FoxH[8][]{%
  \tr{H}^{#2,#3}_{#4,#5}\MeijerM[#1]{#6}{#7}{#8}}
  
\WithSuffix\newcommand\FoxH*[7]{%
  \tr{H}^{#1,#2}_{#3,#4}\MeijerM*{#5}{#6}{#7}}
  
  \DeclarePairedDelimiterX\BivFoxM[7]%
  {\lparen}{\rparen}{#7\delimsize\vert\ \!\!\begin{matrix}\scriptstyle#1 \\ #2\end{matrix}\,|\, \begin{matrix}\scriptstyle#3 \\ #4\end{matrix}\,|\, \begin{matrix}\scriptstyle#5 \\ #6\end{matrix}}  
 
  \WithSuffix\newcommand\BivFoxH*[9]{%
    \tr{H}^{#1}_{#2}\BivFoxM*{#3}{#4}{#5}{#6}{#7}{#8}{#9}}

\begin{document}

\chead{\fontsize{7}{0}\selectfont This work has been submitted to the IEEE for possible publication.  Copyright may be transferred without notice, after which this version may no longer be accessible.}
\cfoot{}
\title{\begin{LARGE}On Multihop Weibull-Fading Communications: Performance Analysis Framework and Applications\end{LARGE}}

\author{%
\begin{normalsize}\mbox{Abdelaziz~Soulimani, Student Member, IEEE, Mustapha~Benjillali, Senior Member,~IEEE,} \mbox{Hatim~Chergui,Member,~IEEE, and Daniel~B.~da~Costa,~Senior Member,~IEEE}\end{normalsize}\\
\thanks{This work has been presented in part at the International Conference on Wireless Networks and Mobile Communications (WINCOM), Marrakech, Morocco, Oct. 2015.}
\thanks{This work has been presented in part at the  International Wireless Communications and Mobile Computing Conference (IWCMC), Valencia, Spain, Jun. 2017.}
\thanks{%
A. Soulimani, M. Benjillali, and H. Chergui are with the National Institute of Telecommunications (INPT), Rabat, Morocco. [e-mails: \{soulimani, benjillali, chergui\}@ieee.org].}
\thanks{D. B. da Costa is with the Department of Computer Engineering, Federal University of Cear\'a, Sobral, CE, Brazil [email: danielbcosta@ieee.org].}%
}%

\maketitle
\renewcommand{\headrulewidth}{0pt}
\thispagestyle{fancy}\setlength{\headheight}{14pt}
\vspace{-1cm}
\begin{abstract}
The paper presents a comprehensive closed-form performance analysis framework for multihop  communications over Weibull fading channels.  
The analyzed scheme consists basically of multiple regenerative relays with generalized high-order quadrature amplitude modulation ($M$-QAM) transmissions.
To take into consideration the channel fading in the mmWave range, we adopt the advocated Weibull model for its flexible ability to cover different channel conditions. 
The end-to-end performance is evaluated in terms of outage probability, bit error probability (BER), symbol error probability (SER), block error rate (BLER), ergodic capacity, and energy efficiency (EE).
For all the metrics, we present exact closed-form expressions along with their asymptotic behavior, some in terms of generalized hypergeometric functions. 
Based on the obtained analytical results, we also present a practical application, we derive two BER- and EE-optimal transmit power allocation strategies, and we discuss the resulting performance gains. 
The exactness of our analysis is illustrated by numerical examples, and assessed via Monte-Carlo simulations for different system and channel parameters. 
Finally, as a secondary contribution, noting the increasing popularity of Fox's $\tr{H}$ and bivariate $\tr{H}$ functions, we provide new and generalized codes for computing these functions which are of practical utility in different contexts.
\end{abstract}

\begin{IEEEkeywords}
5G, BER, BLER, Capacity, Fox~$\tr{H}$-function, IoT, mmWave, Multihop Relaying, Outage Probability, QAM, SER, Weibull Fading.
\end{IEEEkeywords}

\newpage
\section{Introduction}


\IEEEPARstart{C}{ooperative} and multihop communications~\cite{Li2002} had arisen as a common solution to increase the coverage, while preserving high throughput and reliability, with low transmission powers. It is a potential approach to overcome the severe channel conditions that usually impact Millimeter wave (mmWave) signals. On the other hand, the drawbacks of multihop relaying (in terms of increased channel use, coordination overhead, and delay) may be dealt with using optimized transmission parameters and network protocols in a comprehensive and cross-layer framework.%

The double-parameterized Weibull distribution has been shown to model accurately narrow- and large-band small scale fading channels~\cite{Yang2007a,Reig2014}. Moreover, the scale parameter can be modeled as a lognormal distribution. As a result, the Weibull distribution can be used to model mmWave communications: cellular and IoT.

Millimeter wave communications~\cite{Niu2015} (and references therein) have gained great interest as a key enabler technology for the fifth generation (5G) of mobile communication systems. The mmWave band offers large unlicensed bandwidths to answer the huge demand for increased capacity, and higher spectral-efficiency. In addition, it presents interesting anti-interference abilities, together with the new spatial processing techniques allowed by the short wavelengths. While this band was under-utilized in the previous wireless systems---mainly due to practical implementation limitations, cost, and stability---it is nowadays very attractive thanks to cost-effective hardware technologies, and novel directional high-gain antennas. Given their short range and weak penetration over different materials, mmWaves also offer efficient spectrum utilization and secure transmissions.%

Besides the heavy path loss in the mmWave band, large scale blockage is a real challenge too. Even though, these drawbacks can be exploited to increase the efficiency of heterogeneous networks. The main model of the path loss in the mmWave band is the well-known log-normal distribution, whose the parameters were computed using collected data from the measurement campaigns in~\cite{Rappaport2013}. On the other hand, several models are proposed for the blockage effect in the mmWave band. Some of them are summarized in~\mbox{\cite[sec. III]{Andrews2017}}.


There is a significant amount of work on the performance analysis of multihop relaying communications. We only list a few here that are in line with the perspective of this work. For instance, in~\cite{Ikki2007}, the authors have analyzed the performance of multihop relaying systems over Weibull fading channels in terms of bit error ratio (BER) and outage probability. However, the authors considered the amplify-and-forward (AF) strategy, and based their analysis on an approximation of the end-to-end signal-to-noise ratio (SNR) resulting in a lower-bound discussion. A similar analysis of the BER and outage probability over multihop Weibull fading channels was also conducted in~\cite{Wang2015}, but in the context of free-space optical communications with only binary pulse position modulations.\\
Millimeter Wave multihop communications are of interest in access, backhaul networks, and hybrid mode (self-backhauling). In~\cite{Barati2016},  an initial access procedure in mmmWave cellular systems is studied, taking into considerations the directional and omni-directional links and system losses. The authors of~\cite{Garcia2015,Li2017} present and analyze dynamic resource assignment procedures for multihop communications in the mmWave band.

In the context of power allocation optimization, the authors of~\cite{Ikki2011} have presented a dual-hop optimization study of amplify-and-forward relaying systems over Weibull fading channels using multiple antennas. For both amplify-and-forward and decode-and-forward relaying strategies, the authors of~\cite{Randrianantenaina2014} have analyzed the performance of multihop communication systems over Nakagami-$m$ fading, and proposed power allocation schemes maximizing the energy efficiency.

To the best of our knowledge, the performance of regenerative multihop schemes, over the mmWave Weibull-modeled fading channels, remains an open problem, especially in terms of error probabilities with high-order $M$-ary quadrature amplitude modulation (QAM) that is of interest in modern and emerging communication systems. This paper completes and extends our effort in~\cite{Soulimani15,Soulimani2017}.%

In this paper, we propose a comprehensive performance analysis framework of multihop communications over Weibull-fading channels. The paper presents an exhaustive discussion of the major performance analysis metrics, from the exact closed-form expressions, to the asymptotic and more insightful bounds, that are of interest for systems designers and engineers. In addition, the presented framework provides methodological guidelines towards the analysis and optimization of modern systems, where multihop and cooperative communications scheme are natively supported in many use cases. The contributions in this work can be summarized in the following points:
\begin{itemize}
\item We generalize and extend the performance analysis framework in~\cite{Soulimani15,Soulimani2017} to real-valued Weibull shape parameters. This new overview provides a compact and handy survey of the common metrics, all in one place for interested readers from the performance analysis community. All expressions are presented in terms of the single variate and bivariate Fox~$\tr{H}$-functions and the trivariate Meijer G-function. In addition, the asymptotic expressions are simple and offer tractable tools to solve resource allocation and system optimization problems.
\item We highlight the effect of the path loss and the blockage facing mmWave signals.
\item We exploit the obtained results to derive two optimal transmit power allocation strategies. We show that considerable performance gains can be obtained with these allocation schemes, when compared to a uniform allocation for example.
\item We provide and discuss a rich set of numerical results reflecting a wide selection of  applications in 5G and Internet of Things (IoT) eco-systems.
\item Finally, and as a secondary yet important contribution, we implement the Fox~$\tr{H}$ and bivariate~$\tr{H}$-functions in Matlab with generalized contours that are independent of the function parameters. Our codes, unlike a few existing versions that are very dependent on the numerical examples of the context where they were developed, would be of interest and may be readily used by a broader community.
\end{itemize}
 
The remainder of this paper is organized as follows. Section~\ref{sec:sysModel} introduces the proposed system model and the adopted notations. Next, the expressions of the analyzed performance metrics are derived in section~\ref{sec:PAanalytic}. Specifically, exact and asymptotic expressions for the outage probability, BER, SER, BLER, ergodic capacity, and EE are derived. 
Some of the obtained results are used in section~\ref{sec:Opt} to derive optimal power allocation strategies for the multihop schemes.
Numerical examples along with simulation results, using our implementation of the hypergeometric functions, are presented in section~\ref{sec:PAsim}. Section~\ref{sec:Extension} summarizes some possible extensions of this paper and section~\ref{sec:conc} concludes the it. Finally, in appendices we include the new Matlab implementation codes for Fox~$\tr{H}$ and bivariate~$\tr{H}$-functions.

\section{System model}\label{sec:sysModel}


In this work, we consider a cooperative transmission from a source node ($\tr{S}$) to a destination ($\tr{D}$) through $(N-1)$ regenerative\footnote{In general, two main relaying classes are available for cooperative communications, namely, the non-regenerative and the regenerative strategies~\cite{Dohler2010}. In this work, we adopt the popular regenerative ``detect-and-forward'' (DetF) relaying scheme~\cite{Chen2006}, where the relays re-transmit (without decoding and re-encoding) the demodulated binary sequences. The adoption of DetF is motivated by its simplicity, reduced delay, and interesting performance with lower processing complexity and channel state information constraints~\cite{Benjillali08b}.} relay nodes $\tr{R}_i$, $i=1,\ldots,N-1$, as shown in Fig. 1. We assume that all nodes are operating in the half-duplex mode, with the same modulation order $M$, and that all transmissions are orthogonal\footnote{Although this is not the optimal serial transmission protocol, it is adopted to simplify the analysis and avoid all the considerations that are out of the scope of the contribution of this paper.}, e.g., over different time or frequency resources. Each receiving node considers only the previous adjacent transmitter, and the direct link between ($\tr{S}$) and ($\tr{D}$) is not taken into consideration (i.e., no signal is received at the destination directly from the source because of considerable path loss).%
\begin{figure}[ht]
  \psfrag{S}[c][c][2]{$\tr{S}$}
  \psfrag{D}[c][c][2]{$\tr{D}$}
  \psfrag{R1}[][c][2]{$\tr{R}_1$}
  \psfrag{R2}[c][c][2]{$\tr{R}_2$}
  \psfrag{Rn}[c][c][2]{$~\tr{R}_{N-1}$}
  \psfrag{x}[c][r][2.0]{$x$}
  \psfrag{x1}[c][r][2.0]{$\hat{x}_1$}
  \psfrag{x2}[c][r][2.0]{$\hat{x}_2$}
  \psfrag{xn}[c][r][2.0]{$\hat{x}_{N-1}$}
  \psfrag{y}[c][c][2.0]{$y$}
  \psfrag{y1}[c][c][2.0]{$y_1$}
  \psfrag{y2}[c][c][2.0]{$y_2$}
  \psfrag{yn}[c][c][2.0]{$y_{N-1}$}
  \begin{center}
\scalebox{0.35}{\includegraphics{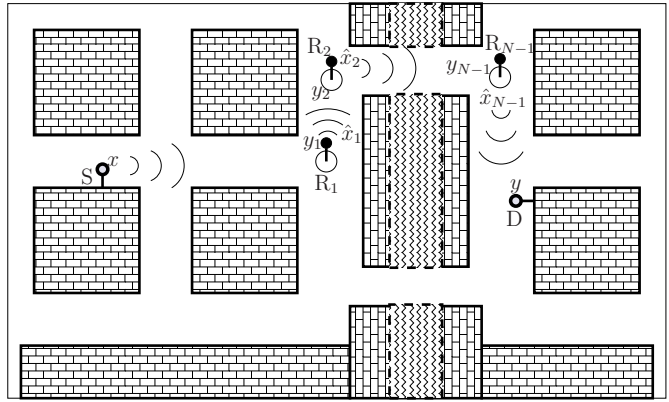}}
\end{center}
  \vspace{-0.8cm}\caption{Adopted system model with $(N{-}1)$ regenerative multihop relays.}\vspace{-0.3cm}	
  \label{fig:SystemModel}
\end{figure}

The transmitted signal from the source node is denoted by $x$, while $y_i$ and $\tilde{x}_i$ are, respectively, the received and transmitted signals at the $i$-th node, and $y$ is the received signal at the destination. Thus, we can express the communication over the $i$-th link, $i=1,\ldots,N$, under the form
\begin{equation}
y_i = \sqrt{\frac{P^\tr{t}_{i-1} B_i}{\mathcal{L}_i}} g_i\tilde{x}_{i-1} + n_i=h_i\tilde{x}_{i-1} + n_i,
\end{equation}
where $\tilde{x}_0=x$, $y_{N}=y$, and $n_i$ denotes the zero-mean additive white Gaussian noise (AWGN) at the $i$-th node, with the same variance $N_0$ over all links. The distance between the source~($S$) and the destination~($D$) is~$d$, and the distance between adjacent nodes is~$d_i$, for $i=1, \ldots, N$. The communication between the nodes $i-1$ and $i$ is subject to a path-loss $\mathcal{L}_i$ given by~\cite{Akdeniz2014}
\begin{equation}\label{Model:PL}
\mathcal{L}_i[\tr{dB}] = \alpha + 10.\beta.\log_{10}(d),
\end{equation}
where $\alpha$ and $\beta$ are given by~\cite[Table 1]{Akdeniz2014} for 28GHz and 73GHz mmWave bands. $B_i$ represents the blockage probability from the 3GPP model~\cite{3GPP36_873} given by
\begin{equation}\label{Model:blockage1}
 B_i = \min\left(\frac{18}{d_i},1\right) \left(1-e^\frac{-d_i}{63}\right) + e^\frac{-d_i}{63},
\end{equation}
for urban areas, and $B_i =\exp\left(\frac{-d_i}{200}\right)$
for suburban areas. Where $d_i$ is expressed in meters.
 
Channel coefficients $g_i$ are assumed to be perfectly known at the receiver, since channel estimation is out of scoup here. The coefficients are also assumed to follow independent but not necessarily identically distributed (i.n.i.d.) flat Weibull fading profiles with parameters $\left( \beta_i,\Omega_i \right)$.

The instantaneous signal-to-noise ratio (SNR) over the $i$-th hop at the receiver may be written as $\gamma_i =|g|_i^2 \overline{\gamma}_i$, where \mbox{$\overline{\gamma}_i=\tr{P}_{i-1}^\tr{t}B_i/N_0/\mathcal{L}_i$} denotes the average SNR at the $i$-th node which is transmitting with a power of $\tr{P}_i^\tr{t}$. 
This SNR is also Weibull distributed~\cite{Simon2005} with parameters $\left(\beta_i/2,\overline{\gamma}_i\Omega_i^2\right)$. Its  probability density function (PDF) is
\begin{equation}\label{eq:WeibullSNR}
\tr{p}_{\gamma_i}(\gamma) =   \frac{\alpha_i}{\phi_i} {\gamma}^{\alpha_i-1} \exp\!\left( \frac{- \gamma^{\alpha_i}}{\phi_i}\right),
\end{equation}
where $\alpha_i=\beta_i/2$ and $\phi_i=\left(\overline{\gamma}_i\Omega_i^2\right)^{\!\alpha_i}$. The cumulative distribution function (CDF) is
\begin{equation}\label{cdfgammak}
	\mathcal{F}_{\gamma_i}(\gamma)= 1 -  \exp\!\left(\!\frac{-\gamma^{\alpha_i}}{\phi_i}\right).
\end{equation}
We introduce also the notation $\overline{\boldsymbol{\gamma}}=[\overline{\gamma}_1, \overline{\gamma}_2,\ldots, \overline{\gamma}_N]$.

\section{Comprehensive Performance Analysis}\label{sec:PAanalytic}



\subsection{Outage Probability}

For the proposed system, an end-to-end outage event occurs when at least one hop go into an outage, namely the transmission rate $\varrho$ of the $i$-th hop is higher than the mutual information over the equivalent channel between the transmitting and receiving nodes. The end-to-end outage probability can be written, for an outage SNR threshold $\gamma_{\tr{th}}=2^{\varrho}-1$, as
\begin{align}
\mathcal{P}_\tr{out} &= \Pr\left[ \min\left(\gamma_1,\dots,\gamma_N\right) \leq \gamma_{\tr{th}} \right],\label{eq:Pout1} \\
					 &= 1 - \Pr\left[ \min\left(\gamma_1,\ldots,\gamma_N\right) \geq \gamma_{\tr{th}} \right] \\
					 &= 1 - \prod_{i=1}^N \left(1-\mathcal{F}_{\gamma_i}(\gamma_{\tr{th}}) \right),\label{eq:Pout2}
\end{align}
Substituting~\eqref{cdfgammak} in~\eqref{eq:Pout2} yields the following compact closed-form expression for the end-to-end outage probability
\begin{equation}
\mathcal{P}_\tr{out} = 1 - \exp\left(- \sum_{i=1}^N \frac{\gamma_{\tr{th}}^{\alpha_i}}{\phi_i}\right). \label{eq:PoutResult}
\end{equation}

For low $\gamma_{\tr{th}}$ values, \eqref{eq:PoutResult} can be approximated by
\begin{equation}
\mathcal{P}_\tr{out} \simeq \sum_{i=1}^N \frac{\gamma_{\tr{th}}^{\alpha_i}}{\phi_i}. \label{eq:PoutAsymptot}
\end{equation}
Assuming identical channel parameters ($\phi_i=\phi, \alpha_i=\alpha$ for all $i=1, \ldots, N$), the outage probability can be further simplified to
\begin{equation}
\mathcal{P}_\tr{out} \simeq \frac{N\gamma_{\tr{th}}^{\alpha}}{\phi}. \label{eq:PoutAsymptotsiml}
\end{equation}
One direct application of this expression, if a given outage probability target $\varPi$ is fixed, would be for example the minimum number of hops between the source and the destination
\begin{equation}
N = \left\lfloor\frac{\phi\varPi}{\gamma_{\tr{th}}^{\alpha}}\right\rfloor,
\end{equation}
where $\lfloor.\rfloor$ denotes the floor function.

\subsection{Bit Error Rate}\label{subsecber}

For regenerative relays, it was shown in~\cite{Morgado2010} that BER can be expressed in terms of the average BER values of each hop
\begin{equation}\label{e2eavgber}
\overline{\mathrm{BER}} = \sum_{i=1}^N \overline{\mathrm{BER}}_i \prod_{j=i+1}^N \left(1-2\overline{\mathrm{BER}}_j \right),
\end{equation}
where $\overline{\mathrm{BER}}_i$ stands for the average BER of the individual $i$-th hop.

\subsubsection*{\textbf{Closed-form exact analysis}}
Over fading channels, $\overline{\mathrm{BER}}_i$ is
\begin{equation}
\overline{\mathrm{BER}}_i = \int_0^{+\infty} \tr{P}_\tr{b}(\gamma)~\!\tr{p}_{\gamma_i}(\gamma)\tr{d}\gamma,
\end{equation}
where, in our case, $\tr{p}_{\gamma_i}(\cdot)$ is given by~\eqref{eq:WeibullSNR}, and $\tr{P}_\tr{b}(\cdot)$ denotes the exact instantaneous BER of an \mbox{$M$-QAM} transmission over a Gaussian channel, which was derived in~\cite{Cho2002} for an arbitrary order $M$ under the form
\begin{equation}
\tr{P}_\tr{b}(\gamma) = \frac{1}{\log_2{\sqrt{M}}} \sum_{m=1}^{\log_2{\sqrt{M}}} \tr{P}_\tr{b}(\gamma,m),
\end{equation}
with
\begin{equation}
\tr{P}_\tr{b}(\gamma,m) = \frac{1}{\sqrt{M}} \sum_{n=1}^{\nu_m} \varPhi_{m,n} \erfc\left( \sqrt{\omega_n \gamma} \right),
\end{equation}
$\nu_m = \left(1-2^{-m}\right)\sqrt{M} -1 $, $\omega_n = \frac{3(2n+1)^2 \log_2M}{2M-2}$, and $\varPhi_{m,n} = (-1)^{\lfloor\frac{n2^{m-1}}{\sqrt{M}}\rfloor} \left(2^{m-1} - \lfloor \frac{n2^{m-1}}{\sqrt{M}} + \frac{1}{2} \rfloor \right)$. Hence, the \mbox{$i$-th} hop BER can be written as
\begin{equation}\label{avgberi}
\overline{\mathrm{BER}}_i = \frac{1}{\sqrt{M}\log_2\sqrt{M}}  \sum_{m=1}^{\log_2{\sqrt{M}}} \sum_{n=0}^{\nu_m} \varPhi_{m,n} \zeta_{n,i},
\end{equation}
where
\begin{align}
\zeta_{n,i} &=\int_0^{+\infty}\erfc{\left( \sqrt{\omega_n\gamma} \right)}\tr{p}_{\gamma_i}(\gamma)\tr{d}\gamma \\
 			&=\frac{\alpha_i}{\phi_i} \int_0^{+\infty}\gamma^{\alpha_i-1}\erfc{\left( \sqrt{\omega_n\gamma} \right)} \exp{\left(\frac{-\gamma^{\alpha_i}}{\phi_i}  \right)} \tr{d}\gamma. \label{zeta}
\end{align}
To evaluate this integral, we express the $\exp(\cdot)$ and $\erfc(\cdot)$  functions, under its generalized Fox~$\tr{H}$-function\footnote{The adoption of generalized functions (like the hypergeometric functions family, Meijer~$\tr{G}$-function, Fox~$\tr{H}$-function) is gaining in popularity both in software computation tools (like Mathematica and Matlab) and among the performance analysis community. We also adopt this efficient and accurate computational approach in the present work.} representation,
\begin{align} \label{erfcFox}
\exp(-x) =  \FoxH*{1}{0}{0}{1}{\text{---}}{(0,1)}{x}\qquad \text{and}\qquad \erfc\!\left( \sqrt{x} \right) &= \frac{1}{\sqrt{\pi}} \FoxH*{2}{0}{1}{2}{(1,1)}{(0,1),(1/2,1)}{x}.
\end{align}
Then using~\cite[Theorem 2.9]{Kilbas2004h} we get
\begin{equation}\label{zetacf}
\mathcal \zeta_{n,i} = \frac{\alpha_i\omega_n^{-\alpha_i}}{\sqrt{\pi}\phi_i} \FoxH*{1}{2}{2}{2}{(1-\alpha_i;\alpha_i),(1/2-\alpha_i;\alpha_i)}{(0,1),(-\alpha_i,\alpha_i)}{\phi_i\omega_n^{-\alpha_i}}.
\end{equation}

By substituting~\eqref{zetacf} in~\eqref{avgberi}, and then in~\eqref{e2eavgber}, we obtain a closed-form expression for $\overline{\mathrm{BER}}$. A general Matlab code to implement Fox~$\tr{H}$-function is provided in Appendix~A.

\subsubsection*{\textbf{Asymptotic analysis}}

The asymptotic analysis\footnote{We note that the asymptotic behavior of the resulting hypergeometric functions may be obtained using a direct expansion~\cite[Th. 1.7 and 1.11]{Kilbas2004h}, but this requires the satisfaction of restricted conditions. Consequently, we adopt a general method using the residue theorem, as explained in~\cite{Chergui15}.} of the BER may be done using residue approach as in~\cite{Chergui15}. Then, \eqref{zetacf} becomes for high $\overline{\gamma}_i$ values
\begin{equation}\label{zetaAs}
\begin{aligned}
\mathcal \zeta_{n,i}& \underset{\overline{\gamma}_i \to \infty}{=} \frac{\omega_n^{-\alpha_i}}{\sqrt{\pi}\phi_i}\Gamma\left(\frac{1}{2}+\alpha_i\right).
\end{aligned}
\end{equation}
By replacing~\eqref{zetaAs} in~\eqref{avgberi}, we get
\begin{equation}\label{avgberiAS}
\overline{\mathrm{BER}}_i \underset{\overline{\gamma}_i \to \infty}{=} \frac{\Gamma\left(\frac{1}{2}+\alpha_i\right)}{\phi_i\sqrt{\pi M}\log_2\sqrt{M}}  \sum_{m=1}^{\log_2{\sqrt{M}}} \sum_{n=0}^{\nu_m} \varPhi_{m,n}\omega_n^{-\alpha_i}.
\end{equation}
One direct application of this result is the diversity order of each hop
\begin{equation}
 \delta_i = -\lim_{\overline{\gamma}_i\to\infty} \frac{\log{\overline{\mathrm{BER}}_i(\overline{\gamma}_i)}}{\log{\overline{\gamma}_i}}= \alpha_i,
\end{equation}
and, using~\eqref{e2eavgber} and~\eqref{avgberi}, we get the diversity order of the end-to-end system  
\begin{equation}\label{berdivord}
\delta_{\tr{e2e}} = \min_{i=1}^N\left(\alpha_i\right).
\end{equation} 

Since $\prod_{j=i+1}^N \left(1-2\overline{\mathrm{BER}}_j \right) \to 1$ for high values of \mbox{$\overline{\gamma}_i$ $(i=1,\dots,N)$}, \eqref{e2eavgber} can be simplified to
\begin{equation}\label{avBERasy}
 \overline{\mathrm{BER}} \underset{\overline{\boldsymbol{\gamma}} \to \infty}{\!\!=} \!\sum_{i=1}^{N}\! \frac{\Gamma(\frac{1}{2}+\alpha_i)}{\sqrt{\!\pi M}\log_2\!\!\sqrt{M}} \!\!\left( \frac{N_0 \mathcal{L}_i}{\tr{P}^t_i B_i \Omega_i^2}\right)^{\!\!\!\alpha_i}\sum_{m=1}^{\log_2\!\sqrt{M}}\!\sum_{n=1}^{\mu_m}\!\frac{\Phi_{m,n}}{\omega_n^{\alpha_i}}.
\end{equation}

%
%

\subsubsection*{\textbf{Outdated CSI}}
Due to the outdated channel state information at the receiver, the fading effect can not be compensated. The resulting signal at the output of the equalizer is
\begin{equation}
 y_i'  = \frac{g_i}{\tilde{g}_i}\tilde{x}_{i-1} + n_i'\,,
\end{equation}
where $g_i$ is the channel coefficient, $\tilde{g}_i$ is the estimated fading which, due to the fast fading, is correlated with the true realization $g_i$. $n_i'$ is the noise after the regulation, its variance $N_i'=1/\overline{\gamma}_i$. Since $|g_i|^2$ and $|\tilde{g}_i|^2$ are also Weibull RVs correlated with a factor of $\rho$. We begging by using the result in~\cite{Yacoub05}, to get the PDF of $G_i =  |g_i|^2/|\tilde{g}_i|^2$ defined as
\begin{align}
\tr{p}_{G_i}(z) &= \int_0^{+\infty}x\tr{p}_{|g_i|^2,|\tilde{g}_i|^2}(zx,x) \,\tr{d}x   \\
      &= \frac{\alpha_i^2}{(1-\rho)\Omega_i^2} z^{\alpha_i-1}\int_0^\infty x^{2\alpha_i-1}\exp\left( \frac{-(z^{\alpha_i}+1)x^{\alpha_i}}{(1-\rho)\Omega_i}  \right) \tr{I}_0\left( \frac{2}{1-\rho}\sqrt{\rho\frac{ z^{\alpha_i}x^{2\alpha_i}}{\Omega_i^2}} \right) \,\tr{d}x\label{eq:BivWeibullSNR} \\
      &=\frac{\alpha_i(1-\rho)z^{\alpha_i-1}(z^{\alpha_i}+1)}{\left(z^{2\alpha_i}+(2-4\rho)z^{\alpha_i}+1\right)^{3/2}}, \label{PDFgi_out}\\
      &=\pi(1-\rho)\alpha_i z^{\alpha_i-1} \BivFoxH*{1,0;1,0;1,0}{1,0;0,1;1,3}{\!(-1;\!1,\!2)}{(\text{---})}{(\text{---})\!}{\!(0,\!1)\!}{\!(\frac{1}{2},\!1)\!}{\!(0,\!1),\!(0,\!1),\!(\frac{1}{2},\!1)\!}{\!z^{\alpha_i}\!\!,\rho z^{\alpha_i}\!}.  
\end{align}

Hence the PDF of the instantaneous SNR of the $i$-th hop \mbox{$\tilde{\gamma}_i =G_i/N_i' = \overline{\gamma}_iG_i$} is 
\begin{align}
\tr{p}_{\tilde{\gamma}_i}(\gamma) =  \frac{\pi(1-\rho)\alpha_i }{\overline{\gamma}_i^{\alpha_i}}\gamma^{\alpha_i-1} \label{PDFcf_out}  \BivFoxH*{1,0;1,0;1,0}{1,0;0,1;1,3}{\!(-1;\!1,\!2)}{(\text{---})}{(\text{---})\!}{\!(0,\!1)\!}{\!(\frac{1}{2},\!1)\!}{\!(0,\!1),\!(0,\!1),\!(\frac{1}{2},\!1)\!}{\!\frac{\gamma^{\alpha_i}}{\overline{\gamma}_i^{\alpha_i}},\frac{\rho \gamma^{\alpha_i}}{\overline{\gamma}_i^{\alpha_i}}}. 
\end{align}
Integrating~\eqref{PDFgi_out} we get the CDF of $\tilde{\gamma}_i$  as
\begin{equation}
 \mathcal{F}_{\tilde{\gamma}_i}(\gamma) =\frac{1}{2}+\frac{1}{2} \frac{\gamma^{\alpha_i}-\overline{\gamma}_i^{\alpha_i}}{\sqrt{\gamma^{2\alpha_i}+(2-4\rho)\overline{\gamma}_i^{\alpha_i}\gamma^{\alpha_i}+\overline{\gamma}_i^{2\alpha_i}}}
\end{equation}

With the same notations as~\eqref{avgberi}, the BER of the $i$-th hop in this case is thus
\begin{align} 
\overline{\mathrm{BER}}_i = \frac{1}{\sqrt{M}\log_2\sqrt{M}}  \sum_{m=1}^{\log_2{\sqrt{M}}} \sum_{n=0}^{\nu_m} \varPhi_{m,n} \tilde\zeta_{n,i},\label{avgberiOut} \\
\tilde\zeta_{n,i} =\int_0^{+\infty}\erfc{\left( \sqrt{\omega_n\gamma} \right)}\tr{p}_{\tilde\gamma_i}\left(\gamma\right) \,\tr{d}\gamma.\label{zeta_out}
\end{align}
Using the short hand notation $\Delta = \overline{\gamma}_i^{-\alpha_i}\omega_n^{-\alpha_i}$ we get~\eqref{zeta_out} in closed form by using~\cite[Eq. 2.1]{Mittal1972integral}
\begin{equation}\label{zetaCF_out}
	\tilde\zeta_{n,i}=\frac{\sqrt{\pi}(1-\rho)}{\left(\alpha_i \Delta\right)^{-1}}\BivFoxH*{2,0;1,0;1,0}{3,1;0,1;1,3}{(1-\alpha_i;\alpha_i,\alpha_i),(\frac{1}{2}-\alpha_i;\alpha_i,\alpha_i),(-1;1,2)}{(-\alpha_i;\alpha_i,\alpha_i)}{(\text{---})}{(0,1)}{(\frac{1}{2},1)}{(0,1),(0,1),(\frac{1}{2},1)}{\Delta,\rho\Delta}.
\end{equation}
\subsubsection*{\textbf{Beamforming}}
Although not the focus of this work, we here discuss the impact of beamforming enhanced directivity, since it is a key enabler for mmWave communications. In this case the channel model should be revisited.

Each node is equipped with a high number of antennas, and it is using $t$ transmitting antennas and $r<t$ receiving ones. At the $i$-th hop, we suppose that the channel matrix $H_i$, Weibull randomly distributed, is perfectly known at the transmitter and the receiver. The singular-value decomposition of the channel matrix is $H_i = r U_i.S_i.V_i^\tr{H}$  , where $U_i$ and $V_i$ are \mbox{$r\,\times\,r$} and  \mbox{$t\,\times\,r$} matrices, $.^\tr{H}$ is the Hermitian transpose symbol, and \mbox{$S_i = \tr{diag}(\sigma_{i,1}, \sigma_{i,2}, ..., \sigma_{i,r})$} where \mbox{$\sigma_{i,1} \geq \sigma_{i,2} \geq ...\geq  \sigma_{i,r} > 0$} are the ordered singular values of $H_i/r$. 

The streamed set of $r$ symbols \mbox{$X_i = [x_{i,1}, x_{i,2},\dots, x_{i,r}]^\tr{T}$} is precoded using $V_i$, where $x_{i,1}, x_{i,2},$ $\dots , x_{i,r}$ are the modulated symbols, and $.^\tr{T}$ is the transpose symbol.  Hence the sent signal at the $i$-th hop is $\overline{X}_i = V _i X_i$.

The receiver gets the signal $\overline{Y}_i =  \sqrt{P^\tr{t}_{i-1} B_i/\mathcal{L}_i}H_i.\overline{X}_i + W_i = \sqrt{P^\tr{t}_{i-1} B_i/\mathcal{L}_i}U_i S_i X_i + W_i$, where $W_i$ is the noise vector at the receiver. Multiplying by $\sqrt{\mathcal{L}_i/P^\tr{t}_{i-1}/B_i}U_i^\tr{H}$ from the left we get as a result \mbox{$\tilde{Y}_i = S_i X_i + \tilde{W}_i$}, namely, the $k$-th signal at the receiver is
\begin{equation}
  \tilde{Y}_{i,k} = \sigma_{i,k} x_{i,k} + w_{i,k},~~ k=1,...,r.
\end{equation}

Thus, the SNR at each receiving antenna can be written as
\begin{equation}
  \gamma_{i,k} = \sigma_{i,k}^2 \overline{\gamma}_{i,k}, ~~ k=1,...,r,
\end{equation}
where $\overline{\gamma}_{i,k}$ is the average SNR for the $i$-th hop at the $k$-th receiving antenna. Assuming large number of antennas, according to the Mar\v{c}enko– Pastur law~\cite{Marvcenko1967}, the PDF of $\gamma_{i,k}$ converges almost surely to
\begin{equation}
 f_{i,k}(x) =  (1-c)^+\delta(x) + \frac{1}{2\pi c s_i^2 x } \sqrt{(b-\frac{x}{\overline{\gamma}_{i,k}})^+(\frac{x}{\overline{\gamma}_{i,k}}-a)^+}, 
\end{equation}
where $\delta(.)$ is the Dirac delta function, $(x)^+ = \max(0,x)$, $c = r/t$, $a = (1-\sqrt{c})^2$,  $b = (1+\sqrt{c})^2$, and $s_i$ is the standard deviation of $H_i$.

Finally, the average BER of the $i$-th hop is
\begin{align}
 \overline{\mathrm{BER}}_i = &\frac{2}{\sqrt{ M}\log_2\sqrt{M}}  \sum_{m=1}^{\log_2{\sqrt{M}}} \sum_{n=0}^{\nu_m} \varPhi_{m,n}\int_0^\infty f_{i,k}(x)\erfc\left({\sqrt{\omega_nx}}\right) \, \tr{d}x \\
  = & \frac{2}{\sqrt{ M}\log_2\sqrt{M}}  \sum_{m=1}^{\log_2{\sqrt{M}}} \sum_{n=0}^{\nu_m}  \frac{\varPhi_{m,n}\sqrt{a'b'}}{2\pi c s_i^2} \mathrm{J}_{a,b,n},
\end{align}
where $a'=a\overline{\gamma}_{i,k}$, $b'=b\overline{\gamma}_{i,k}$, and
\begin{equation}
 \mathrm{J}_{a,b,n} = \int_{a'}^{b'}\frac{\sqrt{(1-x/b')(x/a'-1}}{x} \erfc\left({\sqrt{\omega_nx}}\right) \, \tr{d}x.
\end{equation}
This integral can be evaluated in terms of trivariate Mejer G function by expressing all the functions using their representation through Meijer G; 
\begin{equation}
	\begin{aligned}
		&\qquad\erfc(\sqrt{\omega_nx}) = \frac{1}{\sqrt{\pi}}\MeijerG*{2}{0}{1}{2}{1}{0,1/2}{\omega_nx},  \\
		\sqrt{1-\frac{x}{b'}}= \frac{\sqrt{\pi}}{2}&\MeijerG*{1}{0}{1}{1}{3/2}{0}{\frac{x}{b'}},\qquad
		\text{and}\qquad \sqrt{\frac{x}{a'}-1} = \frac{\sqrt{\pi}}{2}\MeijerG*{0}{1}{1}{1}{3/2}{0}{\frac{x}{a'}}.
	\end{aligned}
\end{equation}
By expressing each Meijer G function by its integral form then interchanging the integrals we finally get
\begin{equation}
	\begin{aligned}
		\mathrm{J}_{a,b,n} = &\frac{\pi}{4} \MultiMeijerG*{0,1;1,0;0,1;2,0}{1,1;1,1;1,1;1,2}{1,\frac{b'}{a'},b' \omega_n}{\begin{matrix} 1 \\ 0 \end{matrix}| \begin{matrix} 3/2 \\ 0 \end{matrix}| \begin{matrix} 3/2 \\ 0 \end{matrix}| \begin{matrix} 1 \\ 0,1/2 \end{matrix}}\\ 
		- &\frac{\pi}{4} \MultiMeijerG*{0,1;1,0;0,1;2,0}{1,1;1,1;1,1;1,2}{\frac{a'}{b'},1,a' \omega_n}{\begin{matrix} 1 \\ 0 \end{matrix}| \begin{matrix} 3/2 \\ 0 \end{matrix}| \begin{matrix} 3/2 \\ 0 \end{matrix}| \begin{matrix} 1 \\ 0,1/2 \end{matrix}}.
	\end{aligned}
\end{equation}

\subsection{Symbol Error Rate}

To alleviate the presentation, we refer the reader to~\cite{Muller08} where the authors have derived the SER of $M$-QAM communications using regenerative relays over Nakagami-$m$ channels. Here, we use the same approach.

\subsubsection*{\textbf{Closed-form exact analysis}}
The calculations of the SER reduce to the following integral\footnote{This is quite obvious considering rectangular QAM constellations, and averaging over the fading distribution.}
\begin{equation}\label{SERdef}
\mathcal{I}_{\tr{SER}_i} = \int\limits_{0}^{+\infty} \tr{Q}\left(A\sqrt{\gamma}\right)\tr{Q}\left(B\sqrt{\gamma}\right)p_{\gamma_i}(\gamma)\tr{d}\gamma,
\end{equation}
where $A$ and $B$ are two positive constant coefficients (note that they cannot be both zero), and $p_{\gamma_i}(.)$ is given by~\eqref{eq:WeibullSNR}. To evaluate~\eqref{SERdef}, two cases need to be differentiated:\vskip 0.2cm
$\bullet$ \underline{$A \times B=0$}: we denote $C=\tr{max}(A,B)$. This case is similar to~\eqref{zeta}, and hence 
\begin{equation}
\begin{aligned}\label{SERcf0}
\mathcal{I}_{\tr{SER}_i}\left(C\right) = \frac{\alpha_i}{4\phi_i\sqrt{\pi}}\left(\frac{2}{C}\right)^{\alpha_i}  \FoxH*{1}{2}{2}{2}{(1-\alpha_i,\alpha_i),(\frac{1}{2}-\alpha_i,\alpha_i)}{(0,1),(-\alpha_i,\alpha_i)}{\frac{1}{\phi_i}\left(\frac{2}{C}\right)^{\alpha_i} }.
\end{aligned}
\end{equation}\vskip 0.2cm
$\bullet$ \underline{$A \times B>0$}: In this case, we replace the $\tr{Q}$-function and the exponential with their Fox~$\tr{H}$-Function representations, and 
using~\cite[eq.~$2.3$]{Mittal1972integral}, we get the closed-form expression of \eqref{SERdef} in terms of the bivariate Fox~$\tr{H}$-Function 
\begin{equation}\label{SERcf1}
\mathcal{I}_{\tr{SER}_i}\left(A,B\right) = \frac{\alpha_i}{4\phi_i\pi}\left(\frac{2}{A}\right)^{\alpha_i} \BivFoxH*{2,0;2,0;1,0}{2,1;1,2;0,1}{(1-\alpha_i;1,\alpha_i),(\frac{1}{2}-\alpha_i;1,\alpha_i)}{(-\alpha_i;1, \alpha_i)}{(1,1)}{(0,1),(\frac{1}{2},1) }{(\text{---})}{(0,1)}{\frac{B}{A},\frac{\left(\frac{2}{A}\right)^{\alpha_i}}{\phi_i}}.
\end{equation}
A general Matlab code to implement the bivariate Fox~$\tr{H}$-Function in Appendix~B.

\subsubsection*{\textbf{Asymptotic Analysis}}
To derive the asymptotic expressions of~\eqref{SERcf0} and~\eqref{SERcf1}, we again reuse the residue method and, after a few mathematical derivations (that we omit here for space limitations), we get
\begin{equation}\label{SERas0}
\mathcal{I}_{\tr{SER}_i}\left(C\right) \underset{\overline{\gamma}_i \to \infty}{=} \frac{1}{4\phi_i\sqrt{\pi}}\Gamma\left(1/2+\alpha_i\right)\left(\frac{2}{C}\right)^{\!\alpha_i},
\end{equation}
and
\begin{equation}\label{SERas1}
\mathcal{I}_{\tr{SER}_i}\!\left(A,B\right) \!\underset{\overline{\gamma}_i \to \infty}{=} \!\frac{-\alpha_i}{8\phi_i\pi^2}\left(\frac{2}{A}\right)^{\alpha_i}\!\!\MeijerG*{2}{2}{3}{3}{\!\alpha_i,1/2+\alpha_i,1}{0,\frac{1}{2},1+\alpha_i}{\!\frac{B}{A}},
\end{equation}
where the Meijer~$\tr{G}$-Function \cite{MeijerGWolfram} is used.

Apart from their simplicity, we emphasize the usefulness of these asymptotic expressions especially in terms of computation time for high-order modulations compared to~\cite[eq. 17]{Muller08} and~\eqref{SERcf1} are time-consuming.

\subsection{Block Error Rate}
In this case we are assuming that the fading channel stays invariant for a duration of a block; $T_\tr{B} = lT_\tr{s}$, where $T_\tr{B}$ is the block time, $T_\tr{s}$ is the symbol time, and $l$ is the number of symbols per block called also channel uses~\mbox{(c.u.)}.

The end-to-end block error rate (BLER), noted $\mathscr{E}$, is defined recursively as
\begin{equation}
\mathscr{E} = \mathcal{E}(\tr{D}) =  \mathcal{E}(\tr{R}_{N-1}) + (1-\mathcal{E}(\tr{R}_{N-1}))\mathscr{E}_N,\qquad \text{and}\qquad \mathcal{E}(\tr{R}_{1}) = \mathscr{E}_1
\end{equation}
where$\mathcal{E}(\tr{R}_{N})$ is the cumulated BLER at the \mbox{$k$-th} node and $\mathscr{E}_i$ is the average individual BLER of the \mbox{$i$-th} hop.

Given this definition we can express the average end-to-end BLER in function of the average individual BLER of each hop only. To do so, we can observe that 
\begin{equation}
 \mathscr{E} = \sum_{i=1}^N \mathscr{E}_i - \sum_{i=2}^N \mathcal{E}(\tr{R}_{i-1}) \mathscr{E}_i.
\end{equation}

By recurrence  it can be shown that 
\begin{align}
 \mathscr{E} = &\sum_{i_1=1}^N \mathscr{E}_i -  \sum_{i_1=1}^N \sum_{i_2=i_1+1}^N \!\!\!\!\mathscr{E}_{i_1}  \mathscr{E}_{i_2} \!+\!  \sum_{i_1=1}^N \sum_{i_2=i_1+1}^N \sum_{i_3=i_2+1}^N \!\!\!\!\mathscr{E}_{i_1}  \mathscr{E}_{i_2}  \mathscr{E}_{i_3} + \ldots  + \nonumber \\ 
 &  + (-1)^{k-1} \sum_{i_1=1}^N \sum_{i_2=i_1+1}^N \cdots \sum_{i_k=i_{k-1}+1}^N \prod_{j=1}^k \mathscr{E}_{i_k} + \ldots + (-1)^{N-1}  \prod_{j=1}^N \mathscr{E}_{i_k},
\end{align}
which can be compacted to
\begin{equation}
 \mathscr{E} = \sum_{i=1}^N (-1)^{i-1}\sum_{j=1}^{\binom{N}{i}} \prod_{k\in\mathcal{S}_{i,j}} \mathscr{E}_k,
\end{equation}
where $\mathcal{S}_{i,j} = \{k_1,k_2,\dots,k_i \}_j ~\tr{such}~ {1\leq k_1<k_2<\dots<k_i\leq N}.$ $\mathcal{S}_{i,j}$ represents the \mbox{$j$-th} set of the $\binom{N}{i}$ sets with $i$ elements chosen from \mbox{$\{1, 2, \dots, N\}$}. 

It remains only to determine the expression of the average  individual BLER of each hop. It is defined~\cite{Mary2016}  for the \mbox{$i$-th} hop as 
\begin{equation} \label{BLERdef}
 \mathscr{E}_i = \int_0^{+\infty} \tr{Q}\left(\frac{\left(C(\gamma)-\varrho \right)}{\sqrt{V(\gamma)/l}} \right) \tr{p}_{\gamma_i}(\gamma) \,\tr{d}\gamma,
\end{equation}
where $\varrho$ is the transmission rate,
\begin{align}
	C(\gamma) = \log_2(1+\gamma), \qquad \text{and}  \qquad  V(\gamma) = \frac{\gamma(\gamma+2)}{(\gamma+1)^2}\left(\log_2e\right)^2.
\end{align}

It is clear that the integral in~\eqref{BLERdef} is very challenging to compute in closed-form. The linear approximation of the term with the \mbox{$\tr{Q}$-function} seems to be very helpful;
\begin{align}\label{qApprox}
\tr{Q}\left(\frac{\left(C(\gamma)-\varrho \right)}{\sqrt{V(\gamma)/l}} \right) \approx
    \begin{cases}
	1 & \gamma\leq\gamma_\texttt{-} \\
	\frac{1}{2}-\lambda\sqrt l (\gamma-\gamma_{\tr{th}}) & \gamma_\texttt{-}<\gamma<\gamma_\texttt{+}\\
	0 & \gamma_\texttt{+}\geq\gamma
    \end{cases},
\end{align}
where $\lambda = \frac{1}{2\pi\sqrt{2^{2\varrho}-1}}$, $\gamma_{\tr{th}} =2^{\varrho}-1$, and $\gamma_\pm = \gamma_{\tr{th}} \pm \frac{1}{2\lambda\sqrt l}$.

Hence the integral~\eqref{BLERdef} becomes
\begin{align}\label{BLERCF}
  \mathscr{E}_i  = \lambda\sqrt l \int_{\gamma_\texttt{-}}^{\gamma_\texttt{+}}\mathcal{F}_{\gamma}(\gamma) \,\tr{d}\gamma =1-\frac{\lambda\sqrt l }{\alpha_i}\phi_i^{\frac{1}{\alpha_i}} \left( \Gamma\left( \frac{1}{\alpha_i},\frac{\gamma_\texttt{-}^{\alpha_i}}{\phi_i} \right) - \Gamma\left( \frac{1}{\alpha_i},\frac{\gamma_\texttt{+}^{\alpha_i}}{\phi_i} \right)\right).
\end{align}

\subsection{Ergodic Capacity} \label{sec:capa}
The ergodic capacity corresponds to the maximum long-term achievable rate averaged over all states of the time-varying channel. In the present context of multihop communication with regenerative relays, the average ergodic capacity can be expressed as
\begin{equation}\label{e2ecapa}
\overline{\mathrm{C}} = \min_{i=1,\ldots, N} {\overline{\mathrm{C}}_i},
\end{equation}
where $\overline{C}_i$ is the bandwidth-normalized average ergodic capacity of the $i$-th hop, given by
\begin{equation}\label{capai}
\overline{\mathrm{C}}_i = \int_0^{+\infty} \!\! \log_{2} (1 + \gamma)~ \tr{p}_{\gamma_i}(\gamma) \, \mathrm{d}\gamma. 
\end{equation}

\subsubsection*{\textbf{Closed-form exact analysis}}
In order to evaluate the integral in~\eqref{capai}, we express the $\exp(\cdot)$~\eqref{erfcFox} and the $\log(\cdot)$ functions through the Fox~$\tr{H}$-function representation,
\begin{equation}\label{logFox}
\log\left( 1 + x \right)=  \FoxH*{1}{2}{2}{2}{(1,1),(1,1)}{(1,1),(0,1)}{x}
\end{equation}
and using~\cite[Theorem 2.9]{Kilbas2004h}, we get~\eqref{capai} in closed-form 
\begin{equation}
\begin{aligned}
   \overline{\mathrm{C}}_i  =& \frac{\alpha_i}{\phi_i\log{2} }\FoxH*{3}{1}{2}{3}{(-\alpha_i,\alpha_i),(1-\alpha_i,\alpha_i)}{(0,1),(-\alpha_i,\alpha_i),(-\alpha_i,\alpha_i)}{\frac{1}{\phi_i}}.
\end{aligned}\label{capaicf}
\end{equation}

\subsubsection*{\textbf{Asymptotic expression}}
For high values of $\overline{\gamma}_i$, recalling that $\phi_i=\left(\overline{\gamma}_i\Omega_i^2\right)^{\!\alpha_i}$, the asymptotic behavior of~\eqref{capaicf} can be performed by using again~\cite[sec. IV]{Chergui15}
\begin{equation}
	\label{capaiAS}
	\overline{\mathrm{C}}_i \underset{\overline{\gamma}_i \to \infty}{=} \frac{\Psi_0\left(1\right)}{\alpha_i\log{2}} +\log_2\left(\phi_i\right)
\end{equation}
where $\Psi_0\left(.\right)$ denotes the digamma function~\cite{Bernardo1976}.


\subsection{Energy Efficiency}\label{subsec:eedef}

\subsubsection*{\textbf{Closed-form exact analysis}}
Without loss of generality, we adopt the definition of energy efficiency (EE) based on the consumption factor metric \cite{Murdock2014}. For the proposed system model, the end-to-end bandwidth normalized EE can be expressed as
\begin{equation}\label{EEinst}
\mathcal{EE}_{\tr{e2e}}= \frac{1}{\tr{P_T}}\log\left( 1+\gamma_{\tr{e2e}} \right),
\end{equation}
where $\gamma_{e2e}=\min_{i=1,\ldots, N}\gamma_i$ denotes the equivalent instantaneous end-to-end SNR. While $\tr{P_T}$ is the total consumed power in the system (circuit and transmission powers) 
$\tr{P_T} = \tr{P_c} + \sum_{i=0}^{N-1} \tr{P}_i^\tr{t}$,
where $\tr{P_c}$ is referring to the circuit power consumed during the transmission ($\tr{P_c^t}$), reception ($\tr{P_c^r}$), modulation ($\tr{P_c^m}$), demodulation ($\tr{P_c^d}$), and in the idle mode ($\tr{P_c^i}$). Thus
\begin{align}
\tr{P_c}  =& \sum_{i=0}^{N-1} \left(\tr{P}_{\tr{c},i}^\tr{t}+ \tr{P}_{\tr{c},i}^\tr{m} \right)+ \sum_{i=0}^{N-1} \left(\tr{P}_{\tr{c},i}^\tr{r} + \tr{P}_{\tr{c},i}^\tr{d} \right)+ \sum_{i=0}^{N} \tr{P}_{\tr{c},i}^\tr{i} \\
=& N\left(\tr{P}_{\tr{c},1}^\tr{t} + \tr{P}_{\tr{c},1}^\tr{r} + \tr{P}_{\tr{c},1}^\tr{m} +\tr{P}_{\tr{c},1}^\tr{d} \right) +\left( N+1\right)\tr{P}_{\tr{c},1}^\tr{i},\label{P_cAssu}
\end{align}
where it is implicitly assumed in~\eqref{P_cAssu} that all nodes have a similar power consumption profile.

The average EE may be obtained directly from~\eqref{EEinst} as
\begin{align} 
\overline{\mathcal{EE}}_{\tr{e2e}}  = \mathbb{E}\left[ \mathcal{EE}_{\tr{e2e}}\right]  =\frac{1}{\tr{P_T}} \int\limits_{0}^{\infty}\left(1-\mathcal{F}_{\gamma_{e2e}}\left( e^x-1\right)\right)\,\tr{d}x,\label{eeD}
\end{align}
where $\mathbb{E}[\cdot]$ stands for the mathematical expectation and 
\begin{align}
 \mathcal{F}_{\gamma_{e2e}}\!\left( \gamma\right)=\tr{Pr}\!\left(\min_{i=1, \ldots, N}\gamma_i \leq \gamma\right) =1 - \prod_{i=1}^N \left( 1-\mathcal{F}_{\gamma_{i}}\left( \gamma\right)\right) =1-\exp\!\left(\!-\sum_{i=1}^N \frac{\gamma^{\alpha_i}}{\phi_i}\right)\!.\label{Fe2eD}
\end{align}

To proceed further with the derivation of $\overline{\mathcal{EE}}_{\tr{e2e}}$, we suppose that $\alpha_i =\alpha$, $i=1, \ldots, N$. Then, taking the simplified form of~\eqref{Fe2eD} into consideration in~\eqref{eeD}, and using the short hand notation $\psi=1/\sum_{i=1}^N \left(1/\phi_i\right)$, we get
\begin{align} 
	\overline{\mathcal{EE}}_{\tr{e2e}} &=\frac{1}{\tr{P_T}}\int\limits_{0}^{\infty}\exp\left( - \frac{\left( e^x-1 \right)^\alpha}{\psi}\right)~\!\tr{d}x=\frac{1}{\tr{P_T}}\int\limits_{0}^{\infty}\frac{1}{v+1}\exp\left( - \frac{v^\alpha}{\psi}\right)\,\tr{d}v \nonumber\\
	&=\frac{1}{\tr{P_T}}\int\limits_{0}^{\infty}\FoxH*{1}{1}{1}{1}{(0,1)}{(0,1)}{v} \FoxH*{1}{0}{0}{1}{\text{---}}{(0,1)}{\frac{v^\alpha}{\psi}}\,\tr{d}v \nonumber\\
	&=\frac{1}{\tr{P_T}}\FoxH*{2}{1}{1}{2}{(0,\alpha)}{(0,1),(0,\alpha)}{\frac{1}{\psi}}.\label{eq:CFEE}
\end{align}

\subsubsection*{\textbf{Asymptotic expression}}
Thanks to the residue approach once again, we can easily obtain an asymptotic expression for the end-to-end EE as
\begin{align} 
\overline{\mathcal{EE}}_{\tr{e2e}} &\approx\frac{1}{\alpha\tr{P_T}} \left(\Psi_0\left(1\right)+\ln\left(\sum_{i=1}^N \frac{1}{\phi_i}\right)\right).
\label{eq:ASEE}
\end{align}

\section{Power Allocation Optimization}\label{sec:Opt}


Now, we exploit some of the obtained asymptotic expressions to derive optimal power allocation (PA) strategies for the adopted scheme. This same methodology can be used in the design of practical multihop configurations from 5G or IoT use cases. To simplify the analysis, we assume that all hops have the same shape parameter, namely $\alpha_i=\alpha, i=1,\dots,N.$

\subsection{\textbf{BER-optimal power allocation}}
As a reference for comparison, we first derive the optimal power allocation strategy minimizing the BER given a total transmit power budget, i.e., the following problem needs to be solved
\begin{equation}
\begin{aligned}
 \underset{P_{0}^\tr{t},P_{1}^\tr{t},\ldots,P_{N-1}^\tr{t}}{\min} \overline{\mathrm{BER}} \quad  \tr{s. t.} \quad \begin{cases}
	 \sum_{i=0}^{N-1} \tr{P}_i^\tr{t} = \tr{P_\text{max}}\\
 \tr{P}_i>0, i=0,\ldots,N-1
\end{cases} 
\end{aligned}
\end{equation}
where $\tr{P_{max}}$ is the maximal transmission power budget of the system. Using result in~\eqref{avBERasy} and the notation $\varphi=\Gamma(1/2+\alpha)/(\sqrt{\pi M}\log_2\sqrt{M})$, we get the Lagrangian cost function as
\begin{equation}
 \begin{aligned} 
 \mathcal{J}_\tr{ber}  = \sum_{i=1}^{N} \varphi \left( \frac{N_0 \mathcal{L}_i}{B_i\tr{P}_{i-1}^\tr{t}  \Omega_i^2}\right)^{\!\alpha} \sum_{m=1}^{\log_2\sqrt{M}}\sum_{n=1}^{\mu_m} \frac{\Phi_{m,n}}{\omega_n^{\alpha}} + \lambda_\tr{ber}\left(  \sum_{i=1}^{N} \tr{P}_{i-1}^\tr{t} - \tr{P_T}\right)\!,
\end{aligned}
\end{equation}
where $\lambda_\tr{ber}$ is the Lagrange multiplier. The $N+1$ Karush-Kuhn-Tucker (KKT) conditions can be expressed as
\begin{equation}
\left\{%
\begin{aligned}
\frac{\partial \mathcal{J}_\tr{ber}}{\partial \tr{P}_{i-1}^\tr{t}} &=-\varphi \left( \frac{N_0 \mathcal{L}_i}{B_i \Omega_i^2}\right)^{\!\alpha} \frac{\left(\tr{P}_{i-1}^\tr{t}\right)^{-\alpha-1}}{\alpha} \times\sum_{m=1}^{\log_2\sqrt{M}}\sum_{n=1}^{\mu_m} \frac{\Phi_{m,n}}{\omega_n^{\alpha}} + \lambda_\tr{ber} = 0\\
\frac{\partial \mathcal{J}_\tr{ber}}{\partial \lambda_\tr{ber}} &=  \sum_{i=1}^{N} \tr{P}_{i-1}^\tr{t} - \tr{P_T} = 0;
\end{aligned}
\right.%
\end{equation}
yielding to a system of equation whose solutions, i.e., the optimal transmit powers, are easily obtained as
\begin{align}
\lambda_\tr{ber} \hfill=  \left( \frac{\tr{P_T}}{\sum_{n=0}^N A_n} \right)^{-\alpha-1} \quad   \text{and} \quad  \tr{P}_{i-1}^\tr{t} = \dfrac{A_i}{\sum_{n=0}^N A_n}  \tr{P_T}, ~ i=1, \ldots, N,
\end{align}
where, for $n=1, \ldots, N$, 
\begin{equation}
	A_n = \left( \!\left( \frac{N_0 \mathcal{L}_n}{B_n\Omega_n} \right)^{\!\alpha}\varphi  \!\sum_{m=1}^{\log_2\sqrt{M}}\sum_{n=1}^{\mu_m} \frac{\Phi_{m,n}}{\omega_n^{\alpha}}\right)^{\!\!1/(\alpha+1)}.
\end{equation}

\subsection{\textbf{EE-optimal power allocation}}
In this subsection we derive the power allocation maximizing the energy efficiency of our system, expressly
\begin{equation}
\begin{aligned}
	\underset{P_{0}^\tr{t},P_{1}^\tr{t},\ldots,P_{N-1}^\tr{t}}{\min} -\overline{\mathcal{EE}}_\tr{e2e}
	\quad  \tr{s. t.} \quad \begin{cases}
	\sum_{i=0}^{N-1} \tr{P}_i^\tr{t} \leq \tr{P_{max}}\\
	\tr{P}_i^\tr{t}>0, i=0,\ldots,N-1
\end{cases} 
\end{aligned}
\end{equation}

The Lagrangian of this problem is 
\begin{equation}
 \mathcal{J}_\tr{ee} = -\overline{\mathcal{EE}}_\tr{e2e}\left( \tr{P}_0^\tr{t},\tr{P}_1^\tr{t},\ldots,\tr{P}_{N-1}^\tr{t} \right) -\lambda_\tr{ee} \left( \sum_{i=0}^{N-1} \tr{P}_i^\tr{t} - \tr{P_{\tr{max}}} \right)
\end{equation}
where $\lambda_\tr{ee}$ is the Lagrange multiplier in this case. The $N+1$ KKT conditions are then 
\begin{equation}
 \left\{
 \begin{aligned}
\frac{\partial \mathcal{J}_\tr{ee}}{\partial \tr{P}_{i-1}} - \lambda_\tr{ee} &=  0\\
\sum_{i=0}^{N-1} \tr{P}_i^\tr{t} - \tr{P_{max}} &=0
\end{aligned}
\right.\quad
\text{or, more explicitly}\quad
 \label{systemEE}
\left\{
\begin{aligned}
\displaystyle{\frac{a_j{\tr{P}_{j-1}^\tr{t}}^{-\alpha-1}}{\sum_{i=1}^N a_i{\tr{P}_{i-1}^\tr{t}}^{-\alpha}}} &= \lambda_\tr{ee}/\mathcal{E}_0, \quad j=1,\ldots,N \\
\sum_{i=1}^N \tr{P}_{i-1}^\tr{t} &= \tr{P_{max}}
\end{aligned}
\right.
\end{equation}
where $\mathcal{E}_0 = 1/\tr{P_T}$, and $a_i=(N_0 \mathcal{L}_i/B_i/\Omega_i^2)^\alpha$ for $i=1,\ldots,N$. The solutions of~\eqref{systemEE} can be easily found as
\begin{equation} \label{chiVal}
 \lambda_\tr{ee} = \frac{\mathcal{E}_0}{\tr{P_{max}}},\qquad  \tr{P}_{k-1}^\tr{t} = \frac{\tr{P_{max}}}{\sum_{i=1}^N \left(a_i/a_k\right)^{1/(\alpha+1)}}, \quad k=1, \ldots, N,
\end{equation}
yielding to the optimal transmit powers maximizing the end-to-end EE.
%

\section{Numerical Results}\label{sec:PAsim}


To assess the accuracy of our theoretical analysis, and illustrate the performance of multi-hop relaying systems in the adopted context of mmWave, we present in this section a few numerical scenarios of interest, and we compare our analytical results to Monte-Carlo simulations.

\subsection{{Setup}}
In order to facilitate the readability of the figures, we follow this same convention in all figures when necessary: solid lines represent the exact analytical results, simulations are represented with markers (only, no lines), and asymptotic expressions correspond to dashed lines. Hence, when the markers are on a line, this should be interpreted as a perfect match between simulation and analytical results.

We took into consideration data realistic parameters from ITU, 3GPP, and FCC to reflect realistic mmWave systems. For instance, we consider the following:\\
- noise power: $-174\tr{dBm/Hz}$,\\
- single antenna element gain: $5\tr{dB}$,\\
- receiver front end loss: $4\tr{dB}$,\\
- roise figure: $5\tr{dB}$.\\
All the studied metrics are plotted in terms of the equivalent isotropically radiated power~(EIRP) since its values are directly related to the SNR; the user devices maximal EIRP is fixed at $23\tr{dBm}$, the base station maximal peak EIRP value can reach up to $85\tr{dBm}$ for high antenna gains ($51\tr{dBi}$), but in general it is limited to $43\tr{dBm}$.


\subsection{{Outage Probability}}
\begin{figure}[!ht]
  \psfrag{ylabel}[c][c][1.13]{$\mathcal{P}_\tr{out}$}
  \psfrag{xlabel}[c][c][1.13]{EIRP [dBm]}
  \psfrag{theo}[l][l][1.13]{Theoretical~\eqref{eq:Pout1}}
  \psfrag{Asympt}[l][l][1.13]{Asymptotic~\eqref{eq:PoutAsymptot}}
  \psfrag{d300N3B001Th000----------------------}[l][l][1.13]{$d$=300\,m, $N$=3, $\tr{B_w}$=1\,MHz, ~~~\!$\gamma_\tr{th}$=0\,dB}
  \psfrag{d300N3B200Th000----------------------}[l][l][1.13]{$d$=300\,m, $N$=3, \underline{$\tr{B_w}$=200\,MHz}, $\gamma_\tr{th}$=0\,dB}
  \psfrag{d600N3B200Th000----------------------}[l][l][1.13]{\underline{$d$=600\,m}, $N$=3, $\tr{B_w}$=200\,MHz, $\gamma_\tr{th}$=0\,dB}
  \psfrag{d600N3B200Th-20----------------------}[l][l][1.1]{$d$=600\,m, $N$=3, $\tr{B_w}$=200\,MHz, \underline{$\gamma_\tr{th}$=-20\,dB}}
  \psfrag{d600N2B200Th-20----------------------}[l][l][1.1]{$d$=600\,m, \underline{$N$=2}, $\tr{B_w}$=200\,MHz, $\gamma_\tr{th}$=-20\,dB}
\begin{center}
\scalebox{0.52}{\includegraphics{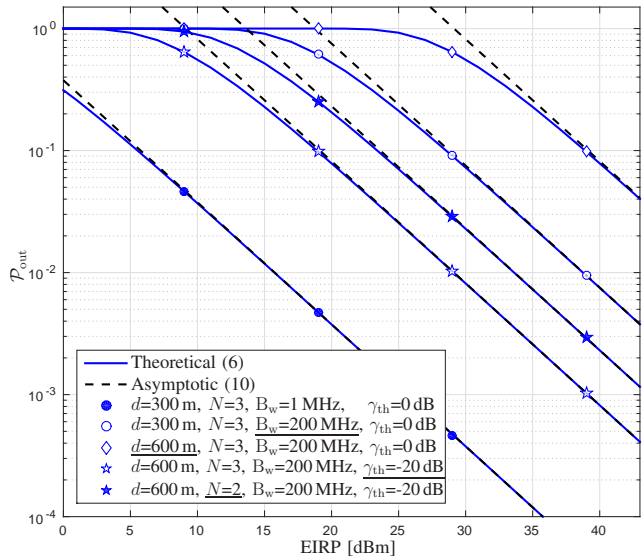}}
\end{center}
\vspace{-0.9cm}
\caption{End-to-end outage probability $\mathcal{P}_\tr{out}$ versus the EIRP per hop of multi-hop Weibull fading channels with similar shape parameters $\beta_i = 2$. In the legend, the underlined parameter is the changed one compared to the previous scenario.}
\vspace{-0.5cm}
\label{fig:Pout}
\end{figure}
Fig.~\ref{fig:Pout} shows the end-to-end outage probability over a three-hop channel. Several scenarios have been studied to highlight some of the system parameters:
\begin{itemize}
	\item The bandwidth has a horizontal shifting effect on the outage probability of the system for the reason that it appears only in the noise power: $N_0\tr{[dBm]} = -174 + 10\log_{10}(\tr{B_w})$. Whereas the same applies to the BER, SER, BLER, and ergodic capacity  metrics. We consider in all the remaining figures, when not specified, that is $\tr{B_w}=200\,\tr{MHz}$. For other $\tr{B_w}$ values (in MHz), the curves can be obtained simply gotten by a horizontally shifting to the right of $10\log_{10}(\tr{B_w}/200)\,\tr{dB}$.
	\item The distance has a double horizontal shifting effect on the outage probability, since it appears in the blockage~\eqref{Model:blockage1} and the path-loss~\eqref{Model:PL} models.
	\item The increasing number of hops seems to decrease the outage probability. This assessment is not always true as it will be discussed later for the end-to-end BER.
	\item In narrow band communications ($\tr{B_w} \propto100$kHz), for IoT applications, the outage probability of the end-to-end system seems to have very low values even with low power nodes.
\end{itemize}
In the figure, it is clear that the implementation of exact closed-form expression~\eqref{eq:PoutResult} matches with simulation results, confirming the exactness of the analysis. We also note the tight asymptotic bound for the region of interest from the outage probability point of view namely $\mathcal{P}_\tr{out}<0.1$.

\subsection{{Capacity}}
\begin{figure*}[h!]
\begin{minipage}[b]{.49\textwidth}
  \psfrag{ylabel}[c][c][1.13]{$\overline{C}/\tr{B_w}$}
  \psfrag{xlabel}[c][c][1.13]{EIRP [dBm]}
  \psfrag{04QAM------------------}  [l][l][1.13]{Theoretical $~4$-QAM}
  \psfrag{64QAM------------------}  [l][l][1.13]{Theoretical $64$-QAM}
  \psfrag{Asympt------------------}  [l][l][1.13]{Asymptotic \eqref{capaiAS}}
  \psfrag{28GHZ------------------}  [l][l][1.13]{Simulation f = $28$ GHz}
  \psfrag{73GHZ------------------}  [l][l][1.13]{Simulation f = $73$ GHz}
  \psfrag{d100}  [l][l][1.13]{\!\!\!\!\!\!\!\!\!\!d = $100$ m}
  \psfrag{d600}  [l][l][1.13]{\!\!\!\!\!\!\!\!\!\!d = $600$ m}
\begin{center}
\scalebox{0.485}{\includegraphics{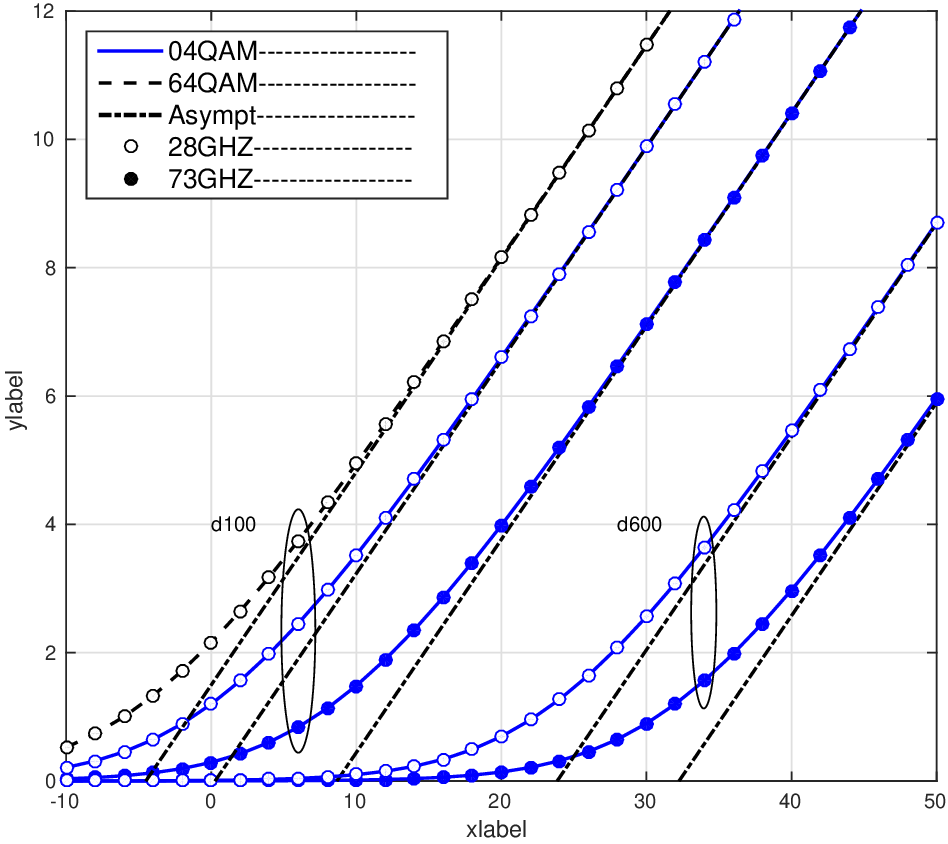}}
\end{center}
\vspace{-0.9cm}
\caption{End-to-end bandwidth-normalized average ergodic capacity $\overline{C}/\tr{B_w}$ versus the EIRP per node for a 3 hops communication system.}
\vspace{-0.5cm}
\label{fig:capa}
\end{minipage}
\hfill
\begin{minipage}[b]{.49\textwidth}
\psfrag{ylabel}[c][c][1.3]{$R(f_1,f_2)$}
  \psfrag{xlabel}[c][c][1.3]{end-to-end distance ($d$)}
  \psfrag{beta1------------------}  [l][l][1.23]{$\beta_i$ = $\sqrt{2}$}
  \psfrag{beta2------------------}  [l][l][1.23]{$\beta_i$ = $(\pi/2+\sqrt{2})/2$}
  \psfrag{beta3------------------}  [l][l][1.23]{$\beta_i$ = $\pi/2$}
  \psfrag{EIRP30}  [l][l][1.3]{EIRP = $30$ dBm}
  \psfrag{EIRP50}  [l][l][1.3]{EIRP = $50$ dBm}
\begin{center}
\scalebox{0.485}{\includegraphics{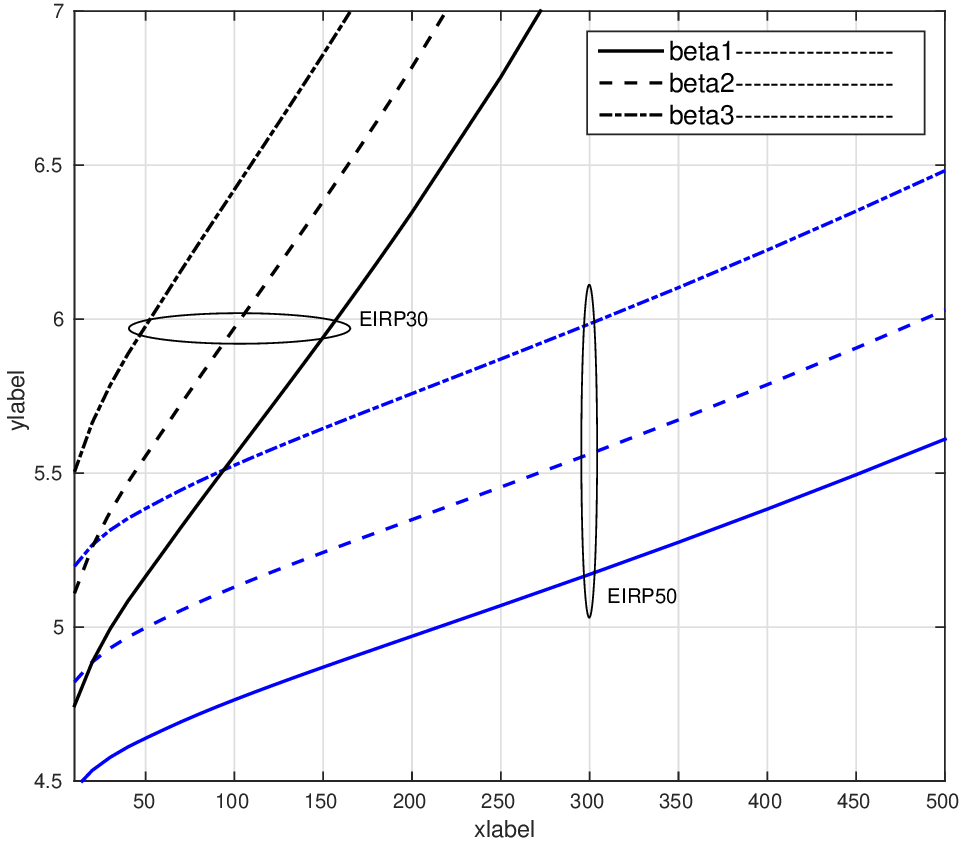}}
\end{center}
\vspace{-0.9cm}
\caption{Ratio of the theoretical end-to-end average ergodic capacities: $R(f_1,f_2)=\overline{C}(28\tr{GHz})/\overline{C}(73\tr{GHz})$ versus the end-to-end distance ($d$).}
\vspace{-0.5cm}
\label{fig:capaRatio}
\end{minipage}
\end{figure*}
Fig.~\ref{fig:capa} depicts the bandwidth-normalized end-to-end ergodic capacity $\overline{C}$ as a function of the EIRP per hop for several scenarios. In this example we are assuming an environment of 3 hops with similar Weibull fading parameters $\beta_i = 2$ and $\Omega_i=1$, $i=1...3$. The figure shows, as it may be expected, that the increasing modulation order increases in creases the end-to-end capacity with a factor related to the difference between the compared modulation orders. Similar to the outage probability, the distance between the end nodes (S,D) has a discernible impact on the ergodic capacity.

To compare the~ capacity of the~ system in the two ~frequency ~bands $f_1=28\,\tr{GHz}$ and $f_2=73\,\tr{GHz}$ for the same bandwidth $\tr{Bw}=200\,\tr{MHz}$, in Fig.~\ref{fig:capaRatio}, we draw the ratio $R(f_1,f_2)$ versus the end to end distance $d$. Here, the communication is established through three hops with the same distance $d_i = d/3$ and similar Weibull parameters in two scenarios; one with an EIRP of $30$ dBm and the other with 50 dBm. 
It is clear that $\overline{C}(f_1)$ is much greater than $\overline{C}(f_2)$ since $R(f_1,f_2)>4$.  We see also that, the curves' slopes are similar within the same scenario, however, $R(f_1,f_2)$ changes significantly given a small variation of the shape parameter (notice that the difference between the values of $\beta_i$ is less than $0.08$ yet the ratio deviates with around $0.5$). The steepness of the curves depends mainly on the SNR represented here with the EIRP. Nonetheless, this comparison is not fair since we use the same bandwidth value for both the capacities, yet, the $73\,\tr{GHz}$ range of frequencies offers more bandwidth and it is suitable for very small distance ranges. Therefore, to decrease $R(f_1,f_2)$, optimal bandwidth values and distances could be chosen.

\subsection{{BER}}
\begin{figure*}[h!t]
\begin{minipage}[b]{.49\textwidth}
 \psfrag{ylabel}[c][c][1.13]{$\overline{\tr{BER}}$}
 \psfrag{xlabel}[c][c][1.13]{EIRP [dBm]}
 \psfrag{Simulation0428}[l][l][1.13]{Simulation $~4$-QAM, $28$ GHz}
 \psfrag{Simulation0473}[l][l][1.13]{Simulation $~4$-QAM, $73$ GHz}
 \psfrag{Simulation6428}[l][l][1.13]{Simulation $64$-QAM, $28$ GHz}
 \psfrag{Simulation1673}[l][l][1.13]{Simulation $16$-QAM, $73$ GHz}
 \psfrag{Asymptot}[l][l][1.13]{Asymptotic~\eqref{avgberiAS}}
 \psfrag{Theoretical}[l][l][1.13]{Theoretical~\eqref{zetacf}}
 \psfrag{D1}[l][l][1.14]{d =   $60$ m}
 \psfrag{D2}[l][l][1.14]{d = $200$ m}
 \psfrag{D3}[l][l][1.14]{d = $600$ m}
\begin{center}
\scalebox{0.485}{\includegraphics{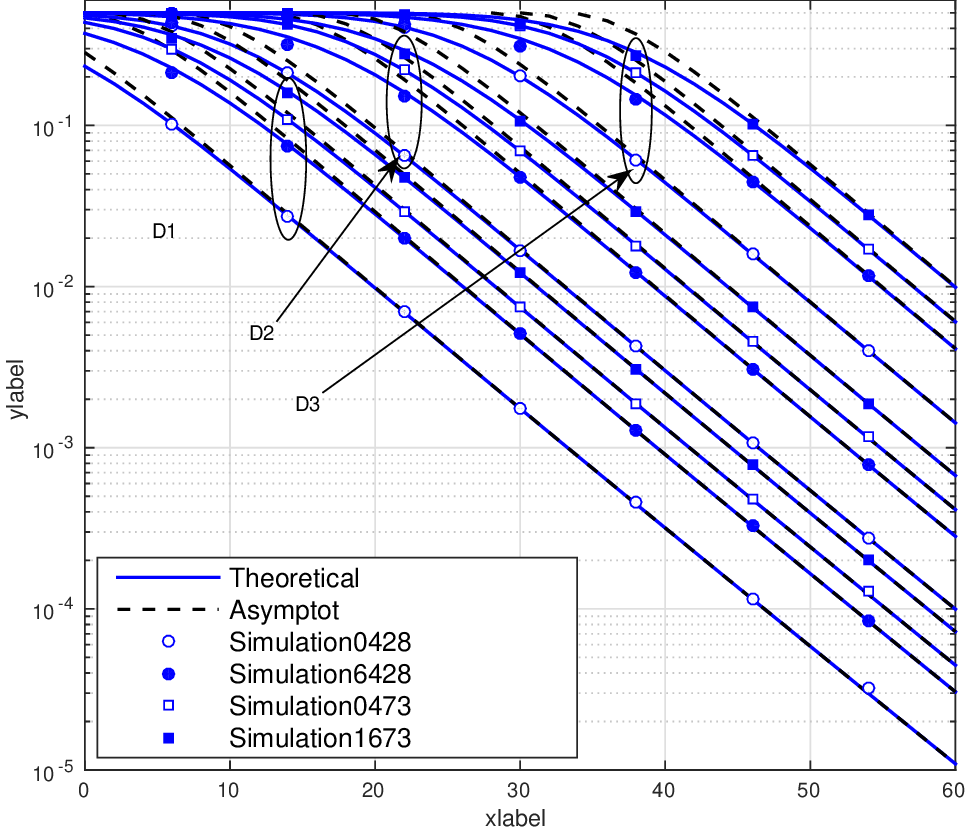}}
\end{center}
\vspace{-0.9cm}
\caption{End-to-end average bit error ratio $\overline{\tr{BER}}$ versus average EIRP per hop. }
\vspace{-0.5cm}
\label{fig:ber}
\end{minipage}
\hfill
\begin{minipage}[b]{.49\textwidth}
 \psfrag{ylabel} [c][c][1.13]{$\overline{\tr{BER}}$}
 \psfrag{xlabel} [c][c][1.13]{Number of hops ($N$)}
 \psfrag{D1----------}[l][l][1.13]{d = ~$800$ m}
 \psfrag{D2----------}[l][l][1.13]{d = $1000$ m}
 \psfrag{EIRP1-}[l][l][1.13]{\!\!\!EIRP1}
 \psfrag{EIRP2-}[l][l][1.13]{EIRP2}
 \psfrag{EIRP1}[l][l][1.13]{EIRP1 = 15 dBm}
 \psfrag{EIRP2}[l][l][1.13]{EIRP2 = 20 dBm}
 \psfrag{EIRP3}[l][l][1.13]{EIRP3 = 30 dBm}
 \psfrag{EIRP4}[l][l][1.13]{EIRP4 = 40 dBm}
 \psfrag{EIRP5}[l][l][1.13]{EIRP5 = 50 dBm}
\begin{center}
\scalebox{0.485}{\includegraphics{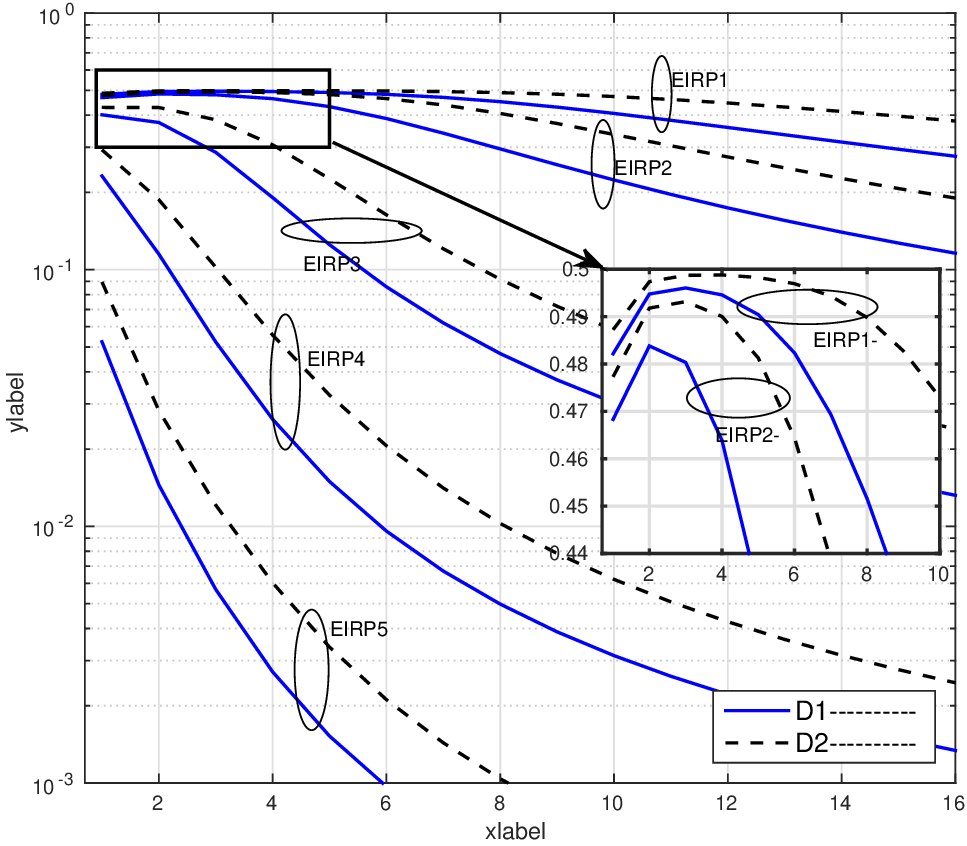}}
\end{center}
\vspace{-0.9cm}
\caption{End-to-end average BER versus the number of hops $N$, with $\beta_i = 2$ and $\Omega_i=1$.}
\vspace{-0.5cm}
\label{fig:ber_hops}
\end{minipage}
\end{figure*}
In Fig.~\ref{fig:ber}, we are plotting, in several scenarios, the end-to-end BER of a three-hops mmWave communication system based on DetF relaying over Weibull fading channels. The figure shows that the modulation order, the end-to-end distance, and the frequency band are key players in designing the system based on the BER metric. For an up-link communication for example, where the device has a maximal EIRP of $23$ dBm, it is very hard to work with high order modulation since the only one bellow $0.1$ for a middler communication link ($200$ meters) is the $4$-QAM in the $28$ GHz band. However, this issue could be overcome using lower bandwidths  (less than the used $200\tr{MHz}$) as it is in general not needed in the up-link or low power-narrow band communications, for many IoT applications, as an example, a $180$kHz bandwidth (the adopted bandwidth value in UL NB-IoT~\cite{3GPPwhite}) would subtract more that $30$dB from the X-axis to get acceptable BER values.

Fig.~\ref{fig:ber_hops} gives insight on the effect of the number of hops on the end-to-end BER. In general the increasing number of hops tends to decrease the end-to-end BER, this fact is true for high SNR values which is the general case of the down-link communications. Concerning the the up-link zone (with less than $23$ dBm for the user devices) the BER tends to increase while increasing the number hops, for instance, two system with $20$ dBm nodes where the first uses a one hop communication and the second uses three hops give approximately the same end-to-end BER.

\begin{figure*}
\begin{minipage}[b]{.49\textwidth}
 \psfrag{ylabel} [c][c][1.13]{$\overline{\tr{BER}}$}
 \psfrag{xlabel} [c][c][1.13]{EIRP (dBm)}
 \psfrag{1Tx1Rx}[l][l][1.13]{$t=r=1$}
 \psfrag{16Tx8Rx}[l][l][1.13]{\!\!\!$t=16,~r=8$}
 \psfrag{f28}[l][l][1.13]{\hskip-1.3cm$f$=28 Ghz}
 \psfrag{f73}[l][l][1.13]{$f$=73 Ghz}
 \psfrag{Theoretical}[l][l][1.13]{Theoretical analysis}
 \psfrag{Perfect1x1trans---------------------}[l][l][1.13]{Perfect CSI with single antennas}
 \psfrag{SimOutdated95}[l][l][1.13]{Simulation, outdated CSI, $\rho=0.95$}
 \psfrag{MIMO16x8}[l][l][1.13]{Perfect CSI with beamforming}
\begin{center}
\scalebox{0.478}{\includegraphics{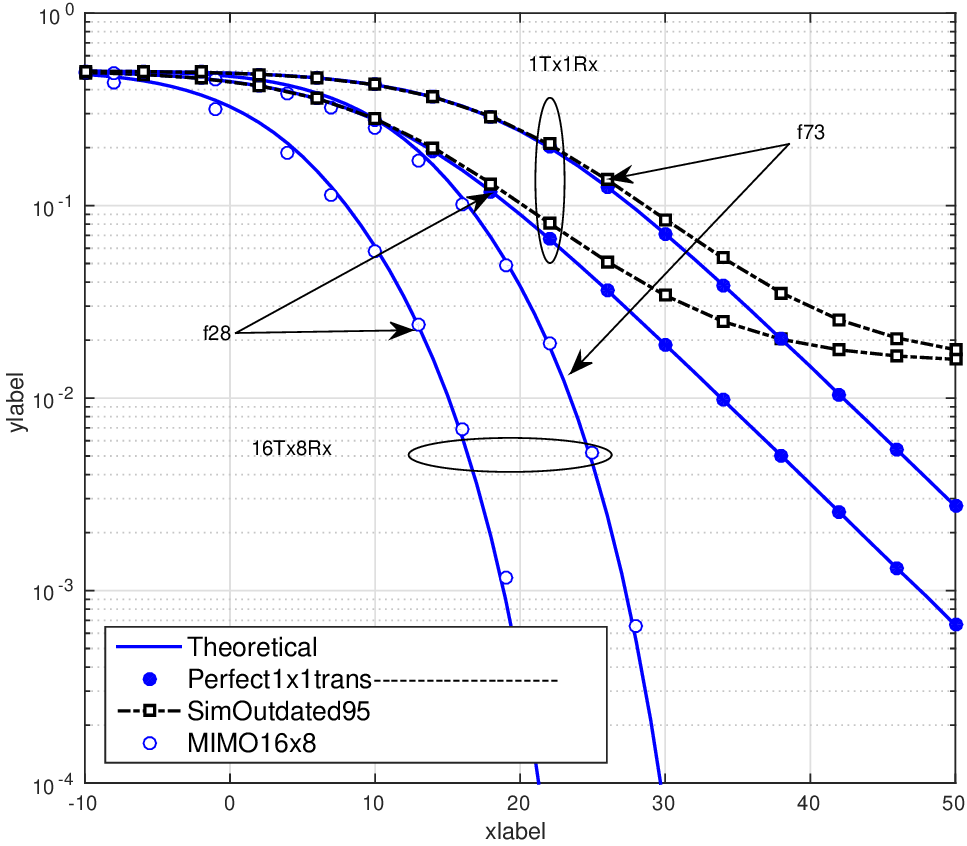}}
\end{center}
\vspace{-0.7cm}
\caption{End-to-end average bit error ratio $\overline{\tr{BER}}$ versus average EIRP per hop for a dual-hop communication, using $16$-QAM modulation highlighting the context of outdated CSI and beamforming.}
\vspace{-0.5cm}
\label{fig:ber_BeamOut}
\end{minipage}
\hfil
\begin{minipage}[b]{.49\textwidth}
  \psfrag{ylabel} [c][c][1.3]{$\overline{\tr{SER}}$}
  \psfrag{xlabel} [c][c][1.3]{EIRP [dBm]}
\begin{center}
\scalebox{0.478}{\includegraphics{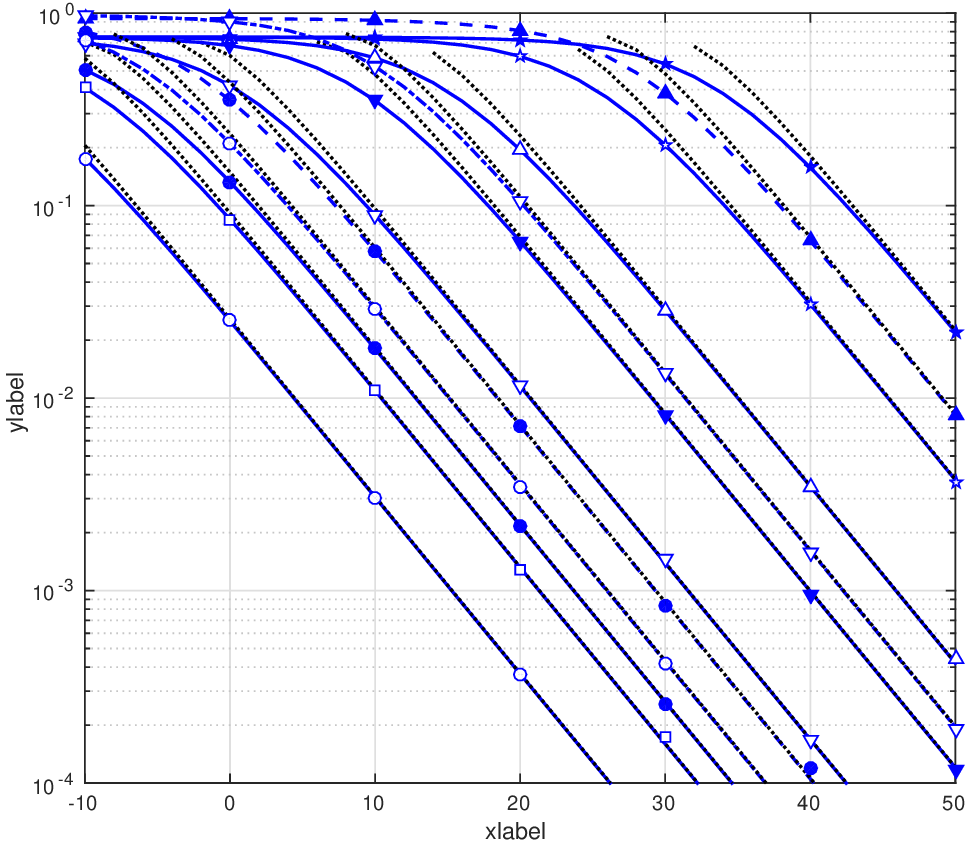}}
\end{center}
\vspace{-0.7cm}
\renewcommand{\arraystretch}{0.6}
\label{table:lagend}
\begin{center}
\begin{scriptsize}
\begin{tabular}{ll || ccl}
 \textcolor{blue}{\rule[1.5pt]{25pt}{0.4pt}}	&	4-QAM 		& &	{$\bullet$} 						&	$d_i = 5$ m	\\
 \textcolor{blue}{\rule[1.5pt]{4pt}{0.4pt}
\rule[1.5pt]{4pt}{0.4pt}
\rule[1.5pt]{4pt}{0.4pt}
\rule[1.5pt]{4pt}{0.4pt}} 				&	16-QAM		& &	{\begin{tiny}$\blacksquare$\end{tiny}} 	&	$d_i =10$ m	\\
 \textcolor{blue}{\rule[1.5pt]{3.5pt}{0.4pt}
\rule[1.5pt]{0.5pt}{0.4pt}
\rule[1.5pt]{3.5pt}{0.4pt}
\rule[1.5pt]{0.5pt}{0.4pt}
\rule[1.5pt]{4pt}{0.4pt}} 				&	64-QAM		& &	{$\blacktriangledown$}				&	$d_i = 30$ m	\\
$\!\cdots\cdots\cdot$						&	Asymptote~\eqref{SERas0}~\eqref{SERas1}	&& 	{$\blacktriangle$}					&	$d_i = 100$ m	\\
\textcolor{blue}{$\bullet$},
\textcolor{blue}{$\blacktriangle$}, ...		&	f = 73 GHz 	& &	{$\bigstar$} 						&	$d_i = 200$ m 	\\
\textcolor{blue}{$\circ$},
\textcolor{blue}{$\vartriangle$}, ...	 		&f = 28 GHz		& &									&				\\
\end{tabular}
\end{scriptsize}
\end{center}
\vspace{-0.7cm}
\caption{End-to-end average symbol error rate $\overline{\tr{SER}}$ versus EIRP per hop for a dual-hop communication.}
\vspace{-0.5cm}
\label{fig:ser}
\end{minipage}
\vspace{-0.5cm}
\end{figure*}

Fig.~\ref{fig:ber_BeamOut} depicts the effect of outdated CSI with correlation factor of $\rho = 0.95$. It is clear from the figure that degraded end-to-end average BER due to the delayed channel state data converge to the same value for higher SNR values, since the effect of additive noise becomes insignificant compared to the outdated CSI. On the obverse side, beamforming has increased the system's performance with the same power allocated to the single antenna transmission especially to effectiveness of beamforming against multipath degradation, not to mention the increased gains at the transmission and reception. Note that we supposed perfect coupling between antennas.

\subsection{{SER}}

In Fig.~\ref{fig:ser}, we present in several scenarios the average end-to-end SER of a dual-hop communication ($\beta_1=1, \beta_2=2$) versus the EIRP per hop of the proposed system:\\
- The results confirm the vulnerability of the mmWave signals to the distance but for small distances, ideal environment of IoT applications, it is very promising. \\
- An other major issue seems to affect this kind of communication is the modulation, it obvious the gap between different modulation orders especially from 4-QAM to 16-QAM which decreases while increasing the modulation order. This is understandable, by the reason of the fact that the $4$-QAM modulation is mainly vulnerable to the AWGN and the phase noise as the boundaries go through the constellation center, while higher modulation are also vulnerable to all the other metrics.\\
- The frequency band has also a perceptible effect similarly to the other metrics.\\
Again, the perfect match between analytical and simulated results, and the tight correspondence between exact and asymptotic analysis, can be appreciated from the figure.

\subsection{{BLER}}

\begin{figure}[h!b]
\begin{minipage}[b]{.49\textwidth}
 \psfrag{ylabel} [c][c][1.2]{$\overline{\tr{BLER}}$}
 \psfrag{xlabel} [c][c][1.2]{EIRP (dBm)}
 \psfrag{Bw1}[l][l][1.2]{$\tr{B_w}=1$MHz, for the }
 \psfrag{RestBw10}[l][l][1.2]{rest $\tr{B_w}=10$MHz.}
 \psfrag{N4}[l][l][1.2]{$N=4$, for the other}
 \psfrag{RestN2}[l][l][1.2]{curves $N=2$.}
 \psfrag{D200}[l][l][1.2]{~~$d=200$}
 \psfrag{Theo04QAM---------------------------}[l][l][1.2]{Theoretical, $4$-QAM}
 \psfrag{Theo16QAM}[l][l][1.2]{Theoretical, $16$-QAM}
 \psfrag{Siml100}[l][l][1.15]{Simulation, $l=100$, and $d=100$m}
 \psfrag{Siml200}[l][l][1.15]{Simulation, $l=200$, and $d=100$m}
 \psfrag{Simd200}[l][l][1.15]{Simulation, $l=100$, and $d=200$m}
\begin{center}
\scalebox{0.475}{\includegraphics{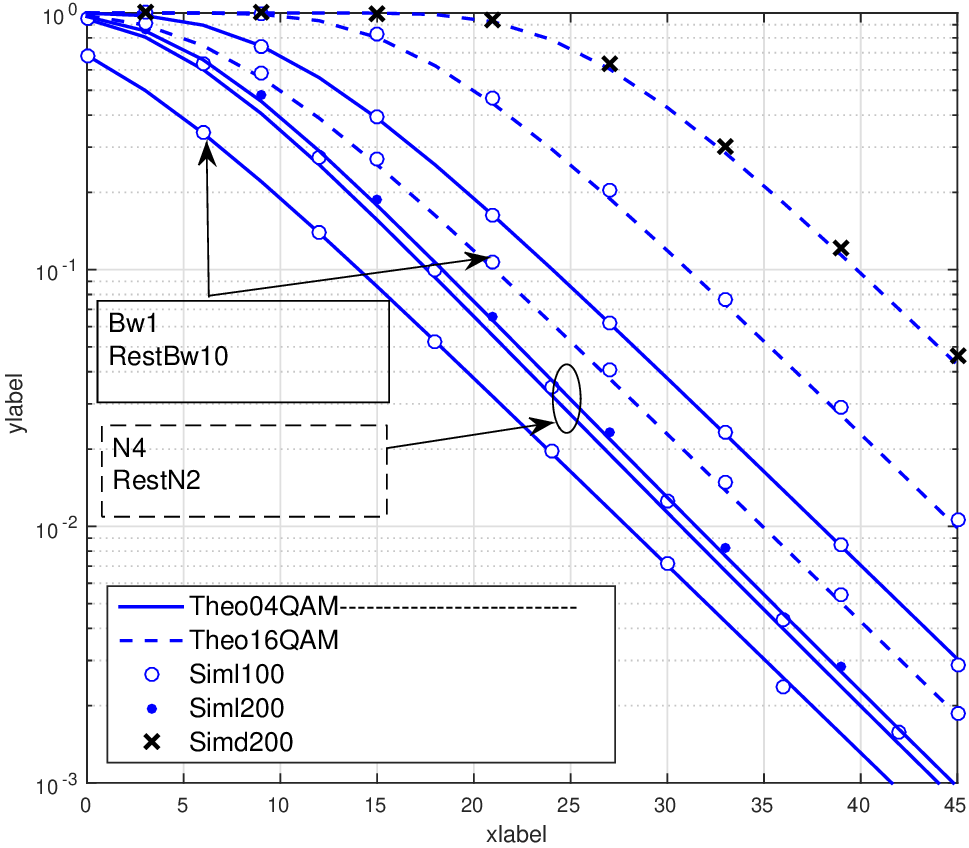}}
\end{center}
\vspace{-0.9cm}
\caption{End-to-end average block error rate $\overline{\tr{BLER}}$ versus average EIRP.}
\vspace{-0.5cm}
\label{fig:bler}
\end{minipage}
\hfil
\begin{minipage}[b]{.49\textwidth}
\psfrag{ylabel}[c][c][1.3]{$\overline{\mathcal{EE}}_\tr{e2e}/\tr{B_w}$}
\psfrag{xlabel}[c][c][1.3]{EIRP [dBm]}
\psfrag{1hop73unif100mBeta1------------}[l][l][1.13]{S0: $d_i=100\tr{m}, f_2, \alpha=0.5$}
\psfrag{1hop73unif100mBeta2----------}[l][l][1.13]{S1: $d_i=100\tr{m}, f_2, \alpha=1$}
\psfrag{2hop73unif100mBeta2}[l][l][1.13]{S2: $d_i=[50~50]\tr{m}, f_2, \alpha=1$}
\psfrag{3hop73unif100mBeta2}[l][l][1.13]{S3: $d_i=[33~33~33]\tr{m}, f_2, \alpha=1$}
\psfrag{20206073unifBeta2}[l][l][1.13]{S4: $d_i=[20~20~60]\tr{m}, f_2, \alpha=1$}
\psfrag{20206073optBeta2}[l][l][1.13]{S5: Optimal power allocation of S4}
\psfrag{4hop28unif200mBeta2}[l][l][1.13]{S6: $d_i=[50~50~50~50]\tr{m}, f_1, \alpha=1$}
\psfrag{4hop28unif200mBeta1}[l][l][1.13]{S7: $d_i=[50~50~50~50]\tr{m}, f_1, \alpha=0.5$}
\begin{center}
\scalebox{0.475}{\includegraphics{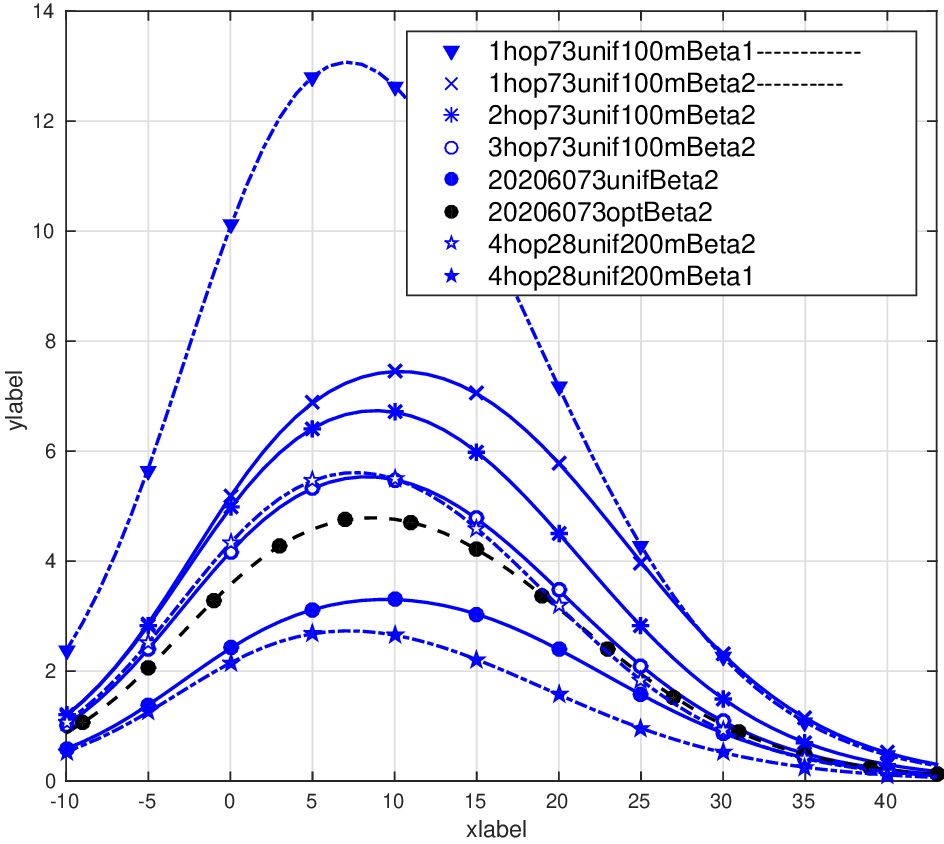}}
\end{center}
\vspace{-0.9cm}
\caption{End-to-end energy efficiency versus EIRP per hop in the bands $f_1 = 28 \tr{GHz}$ and $f_2 = 73 \tr{GHz}$. }
\vspace{-0.5cm}
\label{fig:EE1}
\end{minipage}
\end{figure}

Fig.~\ref{fig:bler} displays the evolution of the BLER in terms of the EIRP. In the figure we highlight some system parameters effect on this metric. First we must note that the approximation of the $Q$-function~\eqref{qApprox} gives good results with an accuracy depending on $\varrho$ and the number of hops. Like the other metrics, the BLER tends to deteriorate when increasing the modulation order. However, high modulation orders have promising results with low distance communications or increasing number of hops.

Large bandwidth communications tend to worsen the BLER due to the accumulated noise through the entire bandwidth. In an other hand, nodes communicating over the $1$~MHz bandwidth show lower BLER values, hence better anti-error performance.

Finally, we used relatively small values of the number of c.u. $l$ to tackle a little bit the ultra-reliable low latency communications (URLLC) as a key enabler of the next cellular networks generation. It seems to have noticeable impact, however minimal, on the end-to-end BLER. 

\subsection{{EE}}

The power inventory to get the total consumed power by the system has been done with the help of results from recent researches~\cite{Marcu2009, Sadhu20177, Kraemer2011, Zou2010, Vidojkovic2012, Wu201764, Fourty2012} on mmWave transceivers. The circuit and the transmit powers  range from a few tens of milliwatts to a few watts depending on the application. The power consumed in the idle mode drops to less than 1\% of the consumed power in the connected mode. In the reminder of this subsection, with the exception of the last figure, we are assuming that the total circuit power consumed in each node equals to $0.5$W.

In Fig.~\ref{fig:EE1}, the end-to-end bandwidth-normalized energy efficiency (BwNEE) is represented in eight scenarios as a function of the average EIRP per hop. The figure highlights the effect of the end-to-end-distance ($d$), adjacent node distances ($d_i$), number of hops ($N$), the frequency band ($f_1$ and $f_2$), and the shape parameter ($\alpha$). The figure shows, from S3 and S4, that distinct distributions of the nodes along this distance affects significantly the BwNEE of the system. Using the proposed power allocation optimization, as illustrated by S4 and S5, the BwNEE has considerably increased. We may also notice, from comparing S0 and S1, that the system becomes less energy efficient when the channel conditions become better (increasing $\alpha$) for low values of the EIRPP (less than $25$dBm) which may be beneficial for the up-link. By studying S1, S2, and S3 we perceive that the increased number of hops decreases system's BwNEE.

\begin{figure}[b]
\begin{minipage}[t]{.49\textwidth}
  \psfrag{ylabel}[c][c][1.3]{number of relays ($N-1$)}
  \psfrag{xlabel}[c][c][1.3]{EIRP [dBm]}
  \psfrag{zlabel}[c][c][1.3]{$\overline{\mathcal{EE}}_\tr{e2e}$/$\tr{B_w}$}
  \psfrag{betapi----------}[l][l][1.3]{$\beta_i=\pi$}
  \psfrag{beta2-----------}[l][l][1.3]{$\beta_i=2$}
\begin{center}
\scalebox{0.47}{\includegraphics{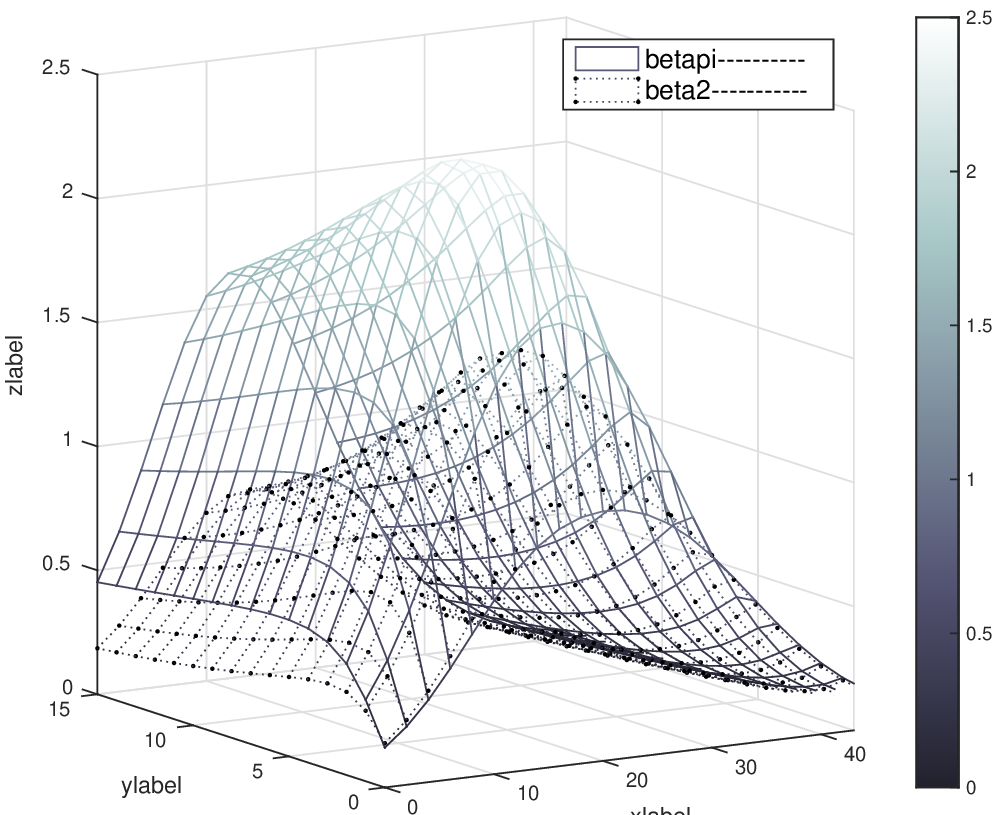}}
\end{center}
\vspace{-0.7cm}
\caption{End-to-end EE $\overline{\mathcal{EE}}_\tr{e2e}$ versus the number of relays and the EIRP per hop. The end-to-end distance of the system is $d=300\,\tr{m}$.}
\label{fig:EE2}
\end{minipage}
\hfil
\begin{minipage}[t]{.49\textwidth}
  \psfrag{ylabel}[c][c][1.3]{$\overline{\mathcal{EE}}_\tr{e2e}$}
  \psfrag{xlabel}[c][c][1.3]{number of relays ($N-1$)}
  \psfrag{De2e010----------}[l][l][1.183]{$d = 10$m}
  \psfrag{De2e100----------}[l][l][1.183]{$d = 100$m}
  \psfrag{De2e400----------}[l][l][1.183]{$d = 400$m}
  \psfrag{Bw180Ptrx100----------}[l][l][1.183]{$\tr{B_w} = 180$kHz, $ \tr{P_c} = 100$mW}
  \psfrag{Bw108Ptrx200----------}[l][l][1.183]{$\tr{B_w} = 180$kHz, $ \tr{P_c} = 200$mW}
  \psfrag{Bw1400Ptrx200--------------}[l][l][1.183]{$\tr{B_w} = 1.4$MHz, \!$\tr{P_c}= 200$mW}
\begin{center}
\scalebox{0.48}{\includegraphics{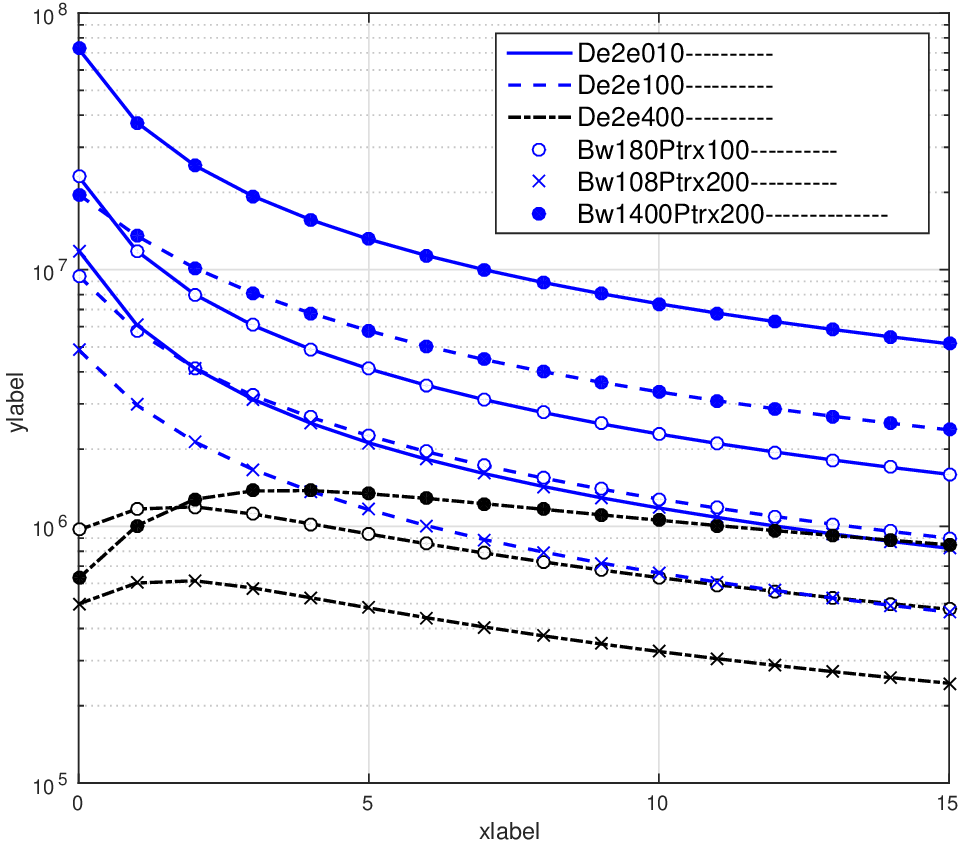}}
\end{center}
\vspace{-0.7cm}
\caption{End-to-end EE $\overline{\mathcal{EE}}_\tr{e2e}$ versus the number of relays ($N$-1). Nodes are communicating in the 28GHz band using an EIRP$ =8\,\tr{dBm}$ with $\beta_i = 2$. }
\label{fig:EE3}
\end{minipage}
\end{figure}
The previous observations about the shape parameter and the number of hops may be misleading, in fact, as shown in Fig.~\ref{fig:EE2}, where we plot the BwNEE versus the number of hops and the EIRP per hop for two values of the shape parameter ($\alpha = \beta_i/2 = 1$ and $\alpha = \beta_i/2 = \pi/2$). The figure shows that the behavior of system's BwNEE depends on all the parameters. This weired behavior of the BwNEE is due to the fact that the transmitting power  has a dual role; an advantage inside the SNR and acts harmfully within the total consumed power of the system.

Finally, in Fig.~\ref{fig:EE3} we studied the energy efficiency of low-power (LP) narrow-band (NB) communications, emulating an IoT environment. We notice that increasing the bandwidth results, in general, in a significant increasing in the energy efficiency of the system. However, in long distance ($400$m) communication very LP~($ \tr{P_c}=100$mW) $180$kHz system (NB-IoT~\cite{3GPPwhite} like scenario) becomes much energy efficient than LP~($ \tr{P_c}=200$mW) 1.4MHz (CAT-M1~\cite{3GPPwhite} like scenario). For low distances, adding relaying nodes seems to be without interest in terms of the EE, it becomes more interesting when the end nodes (S and D) become far away.

\section{Extensions Discussion}\label{sec:Extension}

Several techniques and aspects of the next generation of mobile networks are not taken into consideration or are not investigated in details due to the lack of space, but are part of upcoming extensions of this work. A non exhaustive list of these aspects can be summarized in the following. 


\subsection{Directivity}
Beamforming is one of the main key enablers of the fifth generation. Some aspects that can be investigated to extend this work are:
\begin{itemize}
 \item a more sophisticated analytical analysis of the impact of beamforming,
 \item optimizing the pilot allocation to reduce the effect of pilot contamination,
 \item highlighting the effect of AoA/AoD estimation on the system,
 \item multi-user and access techniques in respect to multiple antenna communication.
\end{itemize}

\subsection{Channel estimation effect}
The channel estimation in a mmWave context represents a real challenge especially when coupled with massive MIMO schemes. Besides outdated CSI, the effect of CSI can be discussed from several other points of view:
\begin{itemize}
 \item limited feedback communication,
 \item erroneous channel estimation,
 \item pilot contamination effect.
\end{itemize}

\subsection{Interference}
As a performance limiting factor, this work can be extended by including the interference aspects into the analysis. This can be done assuming:
\begin{itemize}
 \item full-duplex relays, and analyzing the effect of the residual interference, 
 \item inter node interference, as a result of scheduling scheme,
 \item different recently proposed waveforms.
\end{itemize}

\section{Conclusion}\label{sec:conc}

In this paper, we analyzed and discussed the performance of multihop regenerative relaying in the context of mmWave communications as a key enabler for the next generation of mobile communication systems. Considering a general $M$-QAM modulation order, exact closed-form and asymptotic physical-layer level end-to-end performance metrics (outage probability, ergodic capacity,  BER, BLER, and SER) were derived for Weibull fading links under the form of Fox$\tr{H}$- and Meijer$\tr{G}$-functions based expressions. We also derived expressions for the end-to-end energy efficiency of the analyzed scheme. Based on the obtained results, we computed error and energy efficiency optimal transmit power allocation strategies, and we showed that they offer considerable gains.

Simulation results confirmed the accuracy of our analysis for a large selection of channel and system parameters. As a secondary contribution, we proposed new and generalized implementation of Fox~$\tr{H}$ and bivariate~$\tr{H}$ functions in Matlab.

To complete the analysis, more investigations are necessary, and it will be very interesting to take other aspects into consideration (for example power constraints, delay, channel estimation errors, and transmission scheduling) to get a better, practical, and cross-layer insight into the design and optimization of multihop schemes for the demanding 5G and IoT specifications.

\newpage
\begin{appendices}
\linespread{1.4}
\section{Fox~$\tr{H}$-Function's Matlab Code}
\begin{multicols}{2}
\begin{scriptsize}
    \begin{verbatim}
function out = Fox_H(an, An, ap, Ap,
                  bm, Bm ,bq, Bq,z)
%% Integrand definition
F = @(s)(GammaProd(bm,Bm,s) 
    .* GammaProd(1-an,-An,s).* z.^-s )
    ./ (GammaProd(1-bq,-Bq,s) 
    .* GammaProd(ap,Ap,s));
%% Parameters:
p = length([An Ap]);
q = length([An Ap]);
alphaFox = sum(An)-sum(Ap)+sum(Bm)-sum(Bq);
mu = sum([Bm Bq]) - sum([An Ap]);
betaFox = prod([An Ap].^-[An Ap])
            * prod([Bm Bq].^-[Bm Bq]);
delta = sum([bm bq]) - sum([an ap]) + (p-q)/2;
%% Conditions per contour: 
% Contour L_(c+i*infinity):
condition01=alphaFox>0&&...
         abs(angle(z))<pi*alphaFox/2;
condition02=alphaFox==0&&(delta*mu+...
         real(delta))<-1  && angle(z)==0;
condition0  = condition01 || condition02;
% contour L_(-infinity)
condition11 = (mu>0)&& z~=0;
condition12 = (mu==0) && abs(z)<betaFox ...
         && abs(z)>0;
condition13=(mu==0)&&abs(z)==betaFox...
         &&rea(delta)<-1;
condition1  = condition11||condition12...
         ||condition13;
% contour L_(+infinity)
condition21 = (mu<0)&& z~=0;
condition22 = (mu==0) && abs(z)>betaFox;
condition2  =  condition21 || condition22;
%% Contour preparation:
epsilon = 10^1.2;
Sups = min((1-an)./An); Infs = max(-bm./Bm);
if(isempty(Sups)  && isempty(Infs))
    WPx=1;
elseif(isempty(Sups) && ~isempty(Infs))
    WPx  = Infs +epsilon;
elseif(~isempty(Sups) && isempty(Infs))
    WPx  = Sups -epsilon;
else
    WPx = (Sups + Infs)/2;
end
WayPoints = [WPx-1i*epsilon WPx+1i*epsilon];
%% integration:
if(condition0 || (~condition1 && ~condition2))
    infity = 10;
    out = (1/(2i*pi))*integral(F,WPx-1i*infity,
       WPx+1i*infity);
    return
end
if(~condition1)
    infity = 100;
    if(~isempty(min(-bm./Bm)))
        infity = infity - min(-bm./Bm);
    end
    out = (1/(2i*pi))*integral(F,-infity,...
         -infity, 'Waypoints',WayPoints);
    return
end
if(condition2)
    infity = 100;
    if(~isempty(max((1-an)./An)))
        infity = infity + max((1-an)./An);
    end
    Tol = 10^-5;
    out = (1/(2i*pi))*integral(F,infity...
        ,infity,'Waypoints',WayPoints);
end
%% ***** GammaProd subfunction *****
    function output = GammaProd(p,x,X)
        [pp, XX] = meshgrid(p,X);
        xx = meshgrid(x,X);
        if (isempty(p))
            output = ones(size(X));
        else output = reshape(prod(double(...
             gammaZ(pp+xx.*XX)),2),size(X));
        end
    end
end
\end{verbatim}

\end{scriptsize}
\end{multicols}
\newpage
\section{Bivariate Fox~$\tr{H}$-Function's Matlab Code}
\begin{multicols}{2}
\begin{scriptsize}
\begin{verbatim}
function out = Bivariate_Fox_H(
  an1,alphan1,An1,ap1,alphap1,Ap1,
  bq1,betaq1,Bq1,cn2,Cn2,cp2, Cp2,
  dm2, Dm2, dq2, Dq2, en3, En3,
  ep3, Ep3, fm3, Fm3, fq3, Fq3, x, y)
  %note there is no bm since m=0
%***** Integrand definition *****
F=@(s,t)(GammaProd(1-an1,alphan1,s,An1,t)
  .*GammaProd(dm2,-Dm2, s)
  .* GammaProd(1-cn2,Cn2,s)
  .* GammaProd(fm3,-Fm3,t)
  .* GammaProd(1-en3,En3,t)
  .* (x.^s) .* (y.^t))
  ./ (GammaProd(1-bq1,betaq1,s,Bq1,t)
  .* GammaProd(ap1,-alphap1,s ,-Ap1,t)
  .* GammaProd(1-dq2,Dq2,s)
  .* GammaProd(cp2,-Cp2,s)
  .* GammaProd(1-fq3,Fq3,t)
  .* GammaProd(ep3,-Ep3,t) );
%***** Contour definition *****
% cs
css = 0.1;
Sups = min(dm2./Dm2);
Infs = max((cn2-1)./Cn2);
if(isempty(Sups)  && isempty(Infs))
  cs=1;
elseif(isempty(Sups) && ~isempty(Infs))
  cs  = Infs +css;
elseif(~isempty(Sups) && isempty(Infs))
  cs  = Sups -css;
else
  cs = (Sups + Infs)/2;% Sups< s <Infs
end
% ct
Supt = min(fm3./Fm3);
Inft = max([((-1+an1-alphan1.*cs)./An1)
	    ((en3-1)/En3)]);
if(isempty(Supt)  && isempty(Inft))
  ct=1;
elseif(isempty(Supt) && ~isempty(Inft))
  ct  = Inft +css;
elseif(~isempty(Supt) && isempty(Inft))
  ct = Supt -css;
else
  ct = (5*Supt + Inft)/6;% Supt< t <Inft
end
W = 10; %~infinity
out = real(((1/pi/2i)^2)*quad2d(
      F,cs-1i*W,cs+1i*W,ct-1i*W,ct+1i*W,
      'Singular',true));
%***** GammaProd subfunction *****
  function output = GammaProd(p,x,X,y,Y)
    if(nargin==3)
      [pp, XX] = meshgrid(p,X);
      xx = meshgrid(x,X);
      if (isempty(p))
	output = ones(size(X));
      else
	output = reshape(prod(double(
	gammaZ(pp+xx.*XX)),2),size(X));
      end
    elseif(nargin==5)
      [pp, XX] = meshgrid(p,X);
      xx = meshgrid(x,X);
      yy = meshgrid(y,X);
      [pp, YY] = meshgrid(p,Y);
      if (isempty(p))
	output = ones(size(X));
      else
	output = reshape(prod(double(
	  gammaZ(pp+xx.*XX+yy.*YY)),2),size(X));
      end
    end
  end
end
\end{verbatim}
\end{scriptsize}
\end{multicols}
\begin{small}The gammaZ function  is  available in www.mathworks.com/matlabcentral/fileexchange/3572-gamma
\end{small}
\end{appendices}



\end{document}